\documentclass[onecolumn]{IEEEtran} %

\usepackage[cmex10]{amsmath}
\usepackage{algorithmic}
\usepackage{amssymb}
\usepackage{bm}
\usepackage{stmaryrd}  
\usepackage{amsfonts}
\usepackage{epsfig}

\usepackage{latexsym}
\usepackage{psfrag}
\usepackage{bm}
\usepackage{xspace}
\usepackage{graphicx}

\usepackage{watermark}

\usepackage{amsthm} 
\usepackage{wrapfig}
\usepackage{latexsym}
\usepackage{multicol}
\usepackage{float}
\usepackage{cite}
\usepackage{subfigure}
\usepackage{hyperref} 

\usepackage{etoolbox}
\AtBeginEnvironment{bmatrix}{\setlength{\arraycolsep}{4pt}}

\newcommand{\ignore}[1]{}

\newcommand{\matr}[1]{\mathbf{#1}}
\newcommand{\vect}[1]{\mathbf{#1}}
\newcommand{\code}[1]{\mathcal{#1}}

\newcommand{\codeCQC}[1]{\code{C}_{\mathrm{QC}}^{(r)}}

\newcommand{\defeq}{\triangleq}

\renewcommand{\leq}{\leqslant}
\renewcommand{\geq}{\geqslant}

\def\ill{\dagger}

\newtheorem{lemma}{Lemma}
\newtheorem{theorem}[lemma]{Theorem}

\newtheorem{corollary}[lemma]{Corollary}

\theoremstyle{plain}

\newtheorem{PreDefinition}[lemma]{{\textbf{Definition}}}
  \newenvironment{definition}%
    {\begin{PreDefinition}}{\hfill$\square$\end{PreDefinition}}

\theoremstyle{plain}

\newtheorem{PreRemark}[lemma]{{\textbf{Remark}}}
  \newenvironment{remark}%
    {\begin{PreRemark}\upshape}{\hfill$\square$\end{PreRemark}}

\newtheorem{PreExample}[lemma]{{\textbf{Example}}}
  \newenvironment{example}%
    {\begin{PreExample}\upshape}{\hfill$\square$\end{PreExample}}

\newcommand{\perm}{\operatorname{perm}}

\newcounter{mytempeqcounter}

\usepackage{tikz}
\newcommand*\circled[1]{\tikz[baseline=(char.base)]{
  \node[shape=circle,draw,inner sep=1pt] (char) {#1};}}

\usepackage{xcolor}
\newcommand{\shade}[1]{%
 \colorbox{red!20}{$\displaystyle#1$}}

\title{Bounding the Bethe and the Degree-$M$ Bethe Permanents} 
\author{Roxana Smarandache,~\IEEEmembership{Senior Member,~IEEE,} and Martin Haenggi, \IEEEmembership{Fellow, IEEE}
  \thanks{The support of the NSF (grants DMS 1313221, CCF 1252788 and CCF 1216407) is gratefully acknowledged.}
  \thanks{R.~Smarandache is with the Departments of Mathematics and Electrical Engineering and M.~Haenggi is with the Department of Electrical Engineering, both at the University of Notre Dame, Notre Dame, IN 46556, USA  (e-mail:
    rsmarand@nd.edu, mhaenggi@nd.edu).}
}

\begin{document}
\maketitle

\begin{abstract} In \cite{Vontobel:13}, it was conjectured that the permanent of a $\matr{P}$-lifting $\theta^{\uparrow\matr{P}}$ of a matrix $\theta$ of degree $M$ is  less than or equal to the $M$th power of the permanent $ \perm(\theta)$, i.e., $\perm(\theta^{\uparrow\matr{P}})\leq \perm(\theta)^M$ and, consequently, 
that the degree-$M$ Bethe permanent $\perm_{M,\mathrm{B}} (\theta)$ of a matrix $\theta$ is  less than or equal to the permanent $ \perm(\theta)$ of $\theta$, i.e.,  $\perm_{M, \mathrm{B}} (\theta)\leq  \perm(\theta)$.  In this paper, we prove these related conjectures and show in addition a few properties of the permanent of block matrices that are lifts of a matrix. As a corollary, we obtain an alternative proof of  the inequality $\perm_{\mathrm{B}} (\theta)\leq  \perm(\theta)$ on the Bethe permanent of the base matrix $\theta$ that uses only the combinatorial definition of the Bethe permanent
(the first proof was given by Gurvits in~\cite{2011arXiv1106.2844G}).

  \end{abstract}
\begin{IEEEkeywords}
Bethe free energy, Bethe permanents, permanents, matrix lifts, protographs. 
  \end{IEEEkeywords}

\IEEEpeerreviewmaketitle

\section{Introduction} 
\subsection{Permanents and Bethe permanents}
\IEEEPARstart{T}{he} concept of the {\em Bethe permanent} was introduced 
in \cite{2009arXiv0908.1769H,2011arXiv1108.0065C} 
 to denote the approximation of a permanent of a non-negative matrix\footnote{A non-negative matrix contains only non-negative real entries.} by solving a certain minimization problem of the Bethe free energy with the sum-product algorithm. 
In his paper \cite{Vontobel:13}, Vontobel uses the term {\em Bethe permanent} to denote this approximation and provides reasons why
the approximation works well by showing that the Bethe free energy is a convex function and that the sum-product algorithm finds its minimum efficiently. Although its definition looks simpler than that of the determinant, the permanent does not have the properties of the determinant that enable efficient computation \cite{Minc:78,Minc:87}. 
 Whereas the arithmetic complexity (number of real additions and multiplications) needed to compute the determinant is in $O(n^3)$, Ryser's algorithm, one of the most efficient known algorithms for computing the permanent,  requires $\Theta(n\cdot 2^n)$ arithmetic operations~\cite{Ryser:63}. 
This clearly improves upon the brute-force complexity $O(n \cdot n!) = O(n^{3/2}\cdot (n/e)^n)$ for computing the permanent, but it is still exponential in the matrix size. 
In terms of complexity classes, the computation of the permanent is in the complexity class $\sharp$P~\cite{Valiant:79}, where $\sharp$P is the set of the counting problems associated with the decision problems in the class NP. Even the computation of the permanent of 0-1 matrices restricted to have only three ones per row is $\sharp$P-complete~\cite{Dagum:Luby:Mihail:Vazirani:88}. However,  for circulant matrices, one can exactly calculate the permanent in polynomial time\cite{Codenotti:Shparlinski:Winterhof:02, Metropolis:Stein:Stein:69,Minc:85,Minc:87:1}. 
Later these results were strengthened in various ways in~\cite{Codenotti:Crespi:Resta:97, Codenotti:Resta:01, Codenotti:Resta:02, Bernasconi:Codenotti:Crespi:Resta:99}. 
In contrast to the permanent, the Bethe permanent can be computed efficiently (i.e., in polynomial time).

In the recent paper~\cite{2011arXiv1106.2844G}, Gurvits  shows that  the permanent of a matrix is lower bounded by the Bethe permanent of that matrix, $\perm_{\mathrm{B}}(\theta) \leq \perm(\theta), $  and  discusses conjectures on the constant $C$ in the inequality $\perm(\theta) \leq C \cdot \perm_{\mathrm{B}} (\theta)$. 
 Also, in \cite{ruozzi:12},  Ruozzi showed an analogous result for log-supermodular graphical models. Related to these  results, Vontobel \cite{Vontobel:13} formulates a conjecture that the permanent of an $M$-lift $\theta^{\uparrow\matr{P}}$ of a non-negative matrix $\theta$ is  less than or equal to the $M$th power of the permanent $ \perm(\theta)$, i.e., $\perm(\theta^{\uparrow\matr{P}})\leq \perm(\theta)^M$,  and that the degree $M$-Bethe permanent $\perm_{M,\mathrm{B}} (\theta)$ of a matrix $\theta$ is  less than or equal to the permanent $ \perm(\theta)$ of $\theta$, i.e.,  $\perm_{M, \mathrm{B}} (\theta)\leq  \perm(\theta)$ and proves it  for the particular case of $\theta$ equal to the all-one matrix.  A proof of his general conjecture would imply an alternative proof of  the inequality $\perm_{\mathrm{B}} (\theta)\leq  \perm(\theta)$ that uses only the combinatorial definition of the Bethe permanent.\footnote{The formal definition of the Bethe and $M$-Bethe permanents is given in Definition~\ref{Bethe}.} 

In this paper, we prove this conjecture in its generality. In addition, we prove certain structural properties of the permanent of block matrices that are lifts of a matrix; these matrices are the matrices of interest when studying the degree-$M$ Bethe permanent.

\subsection{Related work} 
The literature on permanents and on adjacent areas (of counting perfect matchings, counting 0-1 matrices with specified row and column sums, etc.) is vast. Apart from the previously mentioned papers, the most relevant papers to our work are the one by Chertkov \& Yedidia \cite{2011arXiv1108.0065C} that studies the so-called fractional free energy functionals and resulting lower and upper bounds on the permanent of a non-negative matrix, the papers~\cite{C:Greenhill:2010}  (on counting perfect matchings in random graph covers),~\cite{Bayati:Nair:06}  (on counting matchings in graphs with the help of the sum-product algorithm\footnote{Computing the permanent is related to counting perfect matchings.}), and~\cite{2009arXiv0908.1769H, Bayati:Shah:Sharma:08,  Sanghavi:Malioutov:Willsky:11} (on max-product/min-sum algorithms based approaches to the maximum weight perfect matching problem). 
Relevant is also the line of work on approximating the permanent of a non-negative matrix using Markov-chain-Monte-Carlo-based methods~\cite{Dagum:Luby:92}, polynomial-time randomized approximation schemes~\cite{Jerrum:Sinclair:Vigoda:04}, 
 and Bethe-approximation based methods or sum-product-algorithm (SPA) based method~\cite{2009arXiv0908.1769H, Yedidia:Freeman:Weiss:05}.\footnote{See~\cite{Vontobel:13} for a more detailed account of these and other related papers.}

\subsection{Paper outline} 
The remainder of the paper is structured as follows. 
 In Section~\ref{sec:definitions}, we list basic notations and definitions, provide the necessary background, and formally define the Bethe permanent and degree-$M$ Bethe permanent.  The following  Section~\ref{lift-perm} contains the results of this paper and consists of three subsections. In Section~\ref{sec:trivial:cover} we show that the bound $\perm(\theta^{\uparrow\matr{P}})\leq \perm(\theta)^M$ is tight, i.e., for every matrix $\theta$, there exists a $\matr{P}$-lifting of $\theta$ for which the bound is satisfied with equality. Sections  \ref{sec:exponent:matrix}--\ref{sec:mapping} present some useful  results on the structure of the permanent of a $\matr{P}$-lifting of degree $M$,  $\theta^{\uparrow\matr{P}}$.
  Section~\ref{sec:coefficient} contains the proof of the conjecture on the permanent of a matrix lifting which follows immediately from these first results,  and Section~\ref{boundingBethe} contains the bounding  results on the Bethe permanent and the degree-$M$ Bethe permanent.  We conclude the paper in Section~\ref{conclusions} and  present a few extra examples of our techniques in the appendix. 

\subsection{Notations and definitions} \label{sec:definitions}
 
Rows and columns of matrices and entries of vectors are indexed starting at $1$.  For an integer $M$, we use the common notation $[M]\defeq \{ 1, \ldots, M\}$. 
 We  use the common notation $h_{ij}$ or $\matr{H}_{ij}$ to denote the $(i,j)$th entry of a matrix $\matr{H}$. 
  For a set $\alpha$, $|\alpha|$ is the cardinality of $\alpha$ (the number of elements in the set $\alpha$). The set of all $M\times M$ permutation matrices is denoted by ${\mathcal P}_M$, and  the set of all permutations on the set $[m]$ is denoted by ${\mathcal S}_m$. 

  \begin{definition}

  Let $\theta = (\theta_{ij})$ be an $m \times m$-matrix over the  integers.
  Its determinant and permanent, respectively,  are defined to be \vspace{-1mm}
  \begin{align*}
    \det(\theta)
      &\defeq \sum_{\sigma \in {\mathcal S}_m}
           \operatorname{sgn}(\sigma)
           \prod_{i\in [m]}
           \theta_{i\sigma(i)} \; ,\quad\quad
    \perm(\theta)
      \defeq \sum_{\sigma \in {\mathcal S}_m}
           \prod_{i\in [m]}
           \theta_{i\sigma(i)} \; , 
  \end{align*}
where $\operatorname{sgn}(\sigma)$ is the signature operator.  

We call the products $\prod_{i\in [m]}
           \theta_{i\sigma(i)}$,  $\sigma \in {\mathcal S}_m$, {\em permanent-products} of $\theta$.  
\end{definition}
  
The following combinatorial description of the Bethe permanent can be found in \cite{Vontobel:13}. We use it here as a definition. 
\begin{definition} \label{Bethe} Let $\theta$ be a non-negative (with non-negative real entries) $m \times m$ matrix and  $M$ be a  positive integer.  
Let $\mathcal{P}_M^{m}$ be the set of all $mM\times mM$ matrices whose blocks are permutation matrices, i.e., 
$$\mathcal{P}_M^{m}\defeq\{\matr{P} = (P_{ij}) \mid P_{ij} \in \mathcal{P}_M,\; \forall i \in[m] \}.$$

For a matrix $\matr{P}\in  \mathcal{P}_M^{m}$, the $\matr{P}$-lifting of $\theta$  is defined as the $mM\times mM $ matrix of weighted permutation matrices:
$$\theta ^{\uparrow\matr{P}}\defeq \begin{bmatrix} \theta_{11}P_{ 11}&\dots & \theta_{1m}P_{1m} \\ \vdots &&\vdots\\ \theta_{m1}P_{m1}&\ldots &\theta_{mm}P_{mm}\end{bmatrix},$$
and the degree-$M$ Bethe permanent of $\theta$ is defined as 
$$ \perm_{\mathrm{B},M}(\theta)\defeq  {\big<\perm(\theta^{\uparrow\matr{P}})\big>
}^{1/M},$$
where the angular brackets $\big<\perm(\theta^{\uparrow\matr{P}})\big>$ represent the arithmetic average of $\perm(\theta^{\uparrow\matr{P}})$ over all $\matr{P} \in \mathcal{P}_M^{m}$. 

The Bethe permanent of $\theta$ is  defined as 
\begin{align*} \perm_{\mathrm{B}}(\theta)\defeq \limsup_{M\rightarrow \infty} \perm_{\mathrm{B},M}(\theta). 
\end{align*}
\vspace*{-0.4cm}
\par
\end{definition}

Since the permanent operator is invariant to the elementary operations of interchanging rows or columns, 
we can assume, when taking the permanent, without loss of generality, that matrices $\matr{P}\in  \mathcal{P}_M^{m}$
have $P_{1j}=P_{i1}=I_M$, for all $i\in [m]$, where $I_M$ is the identity matrix of size $M\times M$. We call such matrices {\em reduced}. 

\begin{definition}
A matrix $\matr{P} =(P_{ij}) \in  \mathcal{P}_M^{m}$ is  {\em reduced} if  $P_{1j}=P_{i1}=I_M$, for all $i,j\in [m]$. The set of all reduced matrices $\matr{P}$  is denoted by $\mathcal{\overline P}_M^{m}$. 
\end{definition}

\begin{remark}  Note that a $\matr{P}$-lifting of a matrix $\theta$ corresponds to an $M$-graph cover of the protograph (base graph) described by $\theta$.  Therefore we can consider  $\theta ^{\uparrow\matr{P}}$ to represent a protograph-based LDPC
code and $\theta$ to be its protomatrix (also called its base matrix or its mother matrix) \cite{thor03}.  \end{remark}

\section{The Permanent of a Matrix-Lift} \label{lift-perm}

In~\cite{Vontobel:13}, it was conjectured that for any  non-negative square matrix $\theta$ and  for any
$\matr{P}\in \mathcal{P}_M^{m}$, 
$$\perm(\theta^{\uparrow\matr{P}})\leq \perm(\theta)^M.$$
In this section we prove this conjecture and several related lemmas on the structure of the $\perm(\theta^{\uparrow\matr{P}})$ of the lift $\theta^{\uparrow\matr{P}}$ of the matrix $\theta$,   for any {non-negative}  matrix $\theta$. 
\subsection{Tightness of the bound} \label{sec:trivial:cover}
We start by showing that there exists at least one lifting for which the bound is tight. The following example shows a lift of the matrix $\theta$ of degree 2 that has maximum permanent $\perm(\theta)^2$. 

\begin{example}\label{trivial}
 Let $\theta$, $\matr{P} \in \mathcal{\overline P}_2^{3}$  and   ${\theta}^{\uparrow\matr{P}}$ as follows: 
  \begin{align*}  \theta\defeq \begin {bmatrix} a&b&c\\d&e&f
\\ g&h&i\end {bmatrix}, \quad \matr{P}  \defeq \begin {bmatrix} I_2&I_2&I_2\\I_2&I_2&I_2
\\ I_2&I_2&I_2\end {bmatrix}, \quad  \setlength{\arraycolsep}{4pt}{\theta}^{\uparrow\matr{P}}\defeq&  \left[\begin {array}{cc|cc|cc} a&0&b&0&c&0\\0&a&0&b&0&c   \\\hline d&0&e&0&f&0\\ 0&d&0&e&0&f\\\hline g&0&h&0&i&0\\ 0&g&0&h&0&i
\end {array}\right]. 
\end{align*}
Equivalently, ${\theta}^{\uparrow\matr{P}}= \theta \otimes \matr{I}$, where $\otimes$ denotes  the Kronecker product of matrices. 
After row and column permutations, which leave the permanent invariant, the matrix ${\theta}^{\uparrow\matr{P}}$ can be rewritten as 
\begin{align*}\setlength{\arraycolsep}{4pt}
 \matr{I}\otimes \theta=   \left[\begin {array}{ccc|ccc} a&b&c&&&\\ d&e&f&&&\\ g&h&i&&&\\\hline &&&a&b&c
\\ &&&d&e&f\\ &&&g&h&i\end {array}\right], \end{align*}
where the empty blocks  contain only zero entries. 
This last matrix is a block-diagonal matrix with permanent equal to the product of the permanents of the matrices on the diagonal, therefore
$$ \perm(\theta^{\uparrow\matr{P}}) = (\perm(\theta))^2. $$
\vspace*{-0.4cm}
\par
\end{example}
 
In fact, a stronger result was shown by Brualdi in~\cite{brualdi:66}. 
 
\begin{theorem} [Theorem 3.1, \cite{brualdi:66}]   Let $\theta$ a non-negative matrix of size $m\times m$ and $P$ a matrix of size $M\times M$.  Then 
$$\perm(\theta\otimes P)=\perm(P \otimes \theta)\geq \perm(\theta)^M\perm(P)^m$$ with equality if and only if  $P$  or $\theta$  has at most one non-zero permanent-product. 
\end{theorem}
Since a permutation matrix  has exactly one non-zero permanent-product, the bound holds with equality when $P$ is a permutation matrix.  The  following corollary follows from this theorem.  
\begin{corollary}  Let $\theta=(\theta_{ij})$ be a non-negative matrix of size $m\times m$ and let $\matr{\tilde P}=(P_{ij})  \in   \mathcal{P}_M^{m}$ such that  $P_{ij}=P$ for all $i,j\in [m]$, where $P$  is a permutation matrix of size $M$.  
Then $$\perm(\theta ^{\uparrow\matr{\tilde P}})= \perm(\theta)^M.$$
\end{corollary}
 
This corollary applies, in particular, for $P=I_M$,  the identity matrix of size $M$.

\subsection{The exponent matrix of a permanent-product}\label{sec:exponent:matrix}

Let $\theta=(\theta_{ij})$ be a non-negative matrix of size $m\times m$ and  $\matr{P}=(P_{ij})\in \mathcal{\overline P}_M^{m}$.  Let 
$\tau \in {\mathcal S}_{mM}$ be a permutation on the set $[mM]$ and let
         $$A_\tau\defeq \prod\limits_{i\in [mM]} (\theta^{\uparrow\matr{P}})_{i\tau(i)}$$ 
 be the permanent-product of $ \theta^{\uparrow\matr{P}}$ pertaining to permutation $\tau$.

\begin{definition}             
           We say that $A_\tau$ is {\em trivially zero} if  there exists $i\in [mM]$ such that $(\theta^{\uparrow\matr{P}})_{i\tau(i)}=0 $ and $ (\theta^{\uparrow\matr{P}})_{i\tau(i)}\neq \theta_{jl}$, for all $j,l\in [m]$,     
    and        
          we say that $A_\tau$ is {\em non-trivially zero} if   $(\theta^{\uparrow\matr{P}})_{i\tau(i)}=  \theta_{jl}$ for some $j,l\in [m]$ and $\theta_{jl}=0$. 
\end{definition} 
In the rest of the paper, all permanent-products considered will be assumed not to be trivially zero. 
Since for each  $i\in [mM]$, there exist $j, l\in [m]$ such that
 $i\in  \left\{(j-1)M+1, \ldots,  jM\right\}$ and $\tau(i)\in \left\{(l-1)M+1, \ldots,  lM\right\}$, 
$(\theta^{\uparrow\matr{P}})_{i\tau(i)}$ is an entry in the weighted permutation matrix $\theta_{jl}P_{jl}$ of $\theta^{\uparrow\matr{P}}$.
By the assumption that $A_\tau$ is not trivially zero, we have
$(\theta^{\uparrow\matr{P}})_{i\tau(i)}=\theta_{jl}$.
Let 
\begin{align}\label{alpha}  \alpha_{ jl}^\tau & \defeq \left\{ i \in \left\{(j-1)M+1, \ldots,  jM\right\} \mid \tau(i)\in \left\{(l-1)M+1,  \ldots,  lM\right\}\right\},\\
r_{ jl}^\tau&\defeq |\alpha_{ jl}^\tau |. 
\end{align} 
 Then,  $(\theta^{\uparrow\matr{P}})_{i\tau(i)}= \theta_{jl}$,  for all $i\in \alpha_{ jl}$ and for all $j,l \in [m]$,  therefore 
 $$\prod\limits_{i=(j-1)M+1}^{jM}  (\theta^{\uparrow\matr{P}})_{i\tau(i)}=\theta_{j1}^{r_{j1}}\theta_{j2}^{r_{j2}}\cdots \theta_{jm}^{r_{jm}}= \prod\limits_{l=1}^{m} \theta_{jl}^{r_{jl}}, \quad\forall j\in[m].$$ 
Since each row and each column of $\theta^{\uparrow\matr{P}}$ must contribute to the product exactly once, the matrix $\alpha_\tau\defeq   (\alpha_{jl}^\tau)_{j,l}$ with its $(j,l)$ entry the set $\alpha_{jl}^\tau$ satisfies
\begin{align}\label{alpha-r} 
&\alpha_{jl}^\tau \bigcap  \alpha_{jl'}^\tau =\emptyset, \;\forall j, l,l'  \in [m], l\neq l',   \quad   \bigcup_{l=1}^m  \alpha_{jl}^\tau =\left\{(j-1)M+1, \ldots,  jM\right\}, 
\end{align}
from which we obtain that  $\sum\limits_{l=1}^{m}r_{jl}^\tau=M$,   for  all $ j\in [m],$ and $\sum\limits_{j=1}^{m}r_{jl}^\tau=M$,  for all $l\in [m].$ 

Therefore, the matrix ${R}_\tau\defeq (r_{ij}^\tau)_{i,j\in[m]}$  corresponding to $A_\tau$ has the property that all its entries are positive, $r_{ij}^\tau\geq 0,$ and  the sums of all entries on each row and each column equal $M$. 
 
We state this fact in the following lemma. 
 
 \begin{lemma} \label{rewriting0} Let $\theta=(\theta_{ij})$ be a non-negative matrix of size $m\times m$  and let $\matr{P}=(P_{ij})\in \mathcal{P}_M^{m}$.  Let 
$\tau \in {\mathcal S}_{mM}$ be a permutation on the set $[mM]$ and let 
         $A_\tau\defeq \prod\limits_{i\in [mM]}
           (\theta^{\uparrow\matr{P}})_{i\tau(i)}$  be a permanent-product of $\theta^{\uparrow\matr{P}}$.  Then, there exists a unique non-negative integer    matrix $R_\tau=(r_{ij}^\tau)$ of size $m\times m$ with the properties
            \begin{align}\label{R-exponents-1}
&\sum\limits_{l=1}^{m}r_{jl}^\tau=M, \quad  \forall j\in [m], \\& \sum\limits_{j=1}^{m}r_{jl}^\tau=M, \quad  \forall l\in [m], \label{R-exponents-2}
\end{align}
              such that
 \begin{align}\label{R-product}&A_\tau=\prod\limits_{j=1}^{m}\prod\limits_{l=1}^{m} (\theta_{jl})^{r_{jl}}.\end{align}

 We call the matrix $R_\tau$ the {\em exponent matrix} of $A_\tau$.  
          \end{lemma}

\subsection{Decomposing the permanent-products of  lifts of matrices } \label{sec:rewriting}
In this subsection, we present a lemma and an algorithm that allows us to rewrite the permanent-products of a $\matr{P}$-lifting of $\theta$ into a form useful for proving the conjecture, namely, as a product of $M$ permanent-products in $\theta$ that are  not necessarily distinct.   
           
\begin{lemma} \label{rewriting} Let $\theta=(\theta_{ij})$ be a non-negative matrix of size $m\times m$  and let $\matr{P}=(P_{ij})\in \mathcal{P}_M^{m}$.  Let 
$\tau \in {\mathcal S}_{mM}$ be a permutation on the set $[mM]$ and let 
         $A_\tau\defeq \prod\limits_{i\in [mM]}
           (\theta^{\uparrow\matr{P}})_{i\tau(i)}$  be a  permanent-product of $\theta^{\uparrow\matr{P}}$.  Then, there exists, not necessarily uniquely,  a set of integers $0\leq t_{\tau\sigma}\leq M$ such that $\sum\limits_{\sigma \in {\mathcal S}_{m}} t_{\tau\sigma}=M$ and 
 \begin{align}\label{product-to-show}A_\tau=
\prod\limits_{\sigma \in {\mathcal S}_{m}} \left(\theta_{1\sigma(1)} \theta_{2\sigma(2)}  \cdots \theta_{m\sigma(m)}\right)^{t_{\tau\sigma}}.
\end{align}
          \end{lemma} 
    
\begin{IEEEproof} 
Let $R_\tau$ be the exponent matrix of $A_\tau$.
For each $\sigma \in {\mathcal S}_{m}$, let $P_{\sigma} \in {\mathcal P}_{m}$ be the $m\times m$ permutation matrix corresponding to $\sigma$ and let $t_{\tau\sigma} \defeq \min\{r_{1\sigma(1)}^\tau, r_{2\sigma(2)}^\tau, \ldots, r_{m\sigma(m)}^\tau\}\geq 0$. Then ${R_\tau}-t_{\tau\sigma} P_{\sigma}$ is a positive matrix with the sums of all entries on each row and each column equal to $M-t_{\tau\sigma}$ and with all its entries  equal to the ones  on the same positions of ${R_\tau}$ except for the entries corresponding to the permutation $\sigma$,  which decreased by the same amount $t_{\tau\sigma}$.  We can index  the set  $\{\sigma \in {\mathcal S}_{m}\}\defeq\{\sigma_{k} \in {\mathcal S}_{m}, k\in [m!]\}$  and compute sequentially 
\begin{align*} R_{\tau,1}&\defeq R_{\tau}, \\ R_{\tau,k+1}&\defeq R_{\tau,k}-t_{\tau{\sigma_k}} P_{\sigma_k}={R_\tau}-\sum_{s=1}^k t_{\tau\sigma_s} P_{\sigma_s}, \; k\geq 2, 
\end{align*}
where the sums of all entries on each row and each column of $R_{\tau,k+1}$ are all equal to $M-\sum\limits_{s=1}^k t_{\tau\sigma_s}.$ Note that after one  round corresponding to a permutation $\sigma$, the entries are either  the same if $t_{\tau\sigma}=0$ or,   if $t_{\tau\sigma}\neq 0$, at least one non-zero entry  in the matrix $R_{\tau,k}$ (corresponding to $t_{\tau\sigma}$) gets changed to a zero entry  in the matrix $R_{\tau,k+1}$ and all the other entry values on the positions corresponding to the permutation $\sigma_k$  decrease by the same amount  $t_{\tau\sigma_k}$. The algorithm runs until all non-zero entries get changed into  zero entries, see Example~\ref{matrixR} for an illustration of this process.   
Consequently,  the matrix $R-\sum\limits_{\sigma \in {\mathcal S}_{m}} t_{\tau\sigma} P_\sigma=0$. This yields  $R=\sum\limits_{\sigma \in {\mathcal S}_{m}} t_{\tau\sigma} P_\sigma,$ leading to 
 $$A_\tau=\prod_{i\in [mM]}  (\theta^{\uparrow\matr{P}})_{i\tau(i)}=\prod\limits_{j=1}^{m}\prod\limits_{l=1}^{m} (\theta_{jl})^{r_{jl}}=
\prod\limits_{\sigma \in {\mathcal S}_{m}} \left(\theta_{1\sigma(1)} \theta_{2\sigma(2)}  \cdots \theta_{m\sigma(m)}\right)^{t_{\tau\sigma}} 
$$
and $\sum\limits_{\sigma \in {\mathcal S}_{m}} t_{\tau\sigma}=M$.

Note that this described decomposition  always works, i.e., the steps presented above can be always performed until all the entries are changed into zero entries. This is due to the Birkhoff-von Neumann theorem on the decomposition of doubly stochastic matrices into a convex combination of permutation matrices that insures that the doubly stochastic matrix $\frac{1}{M} R_\tau$  can be decomposed indeed as a convex sum of permutation matrices.\footnote{A matrix is doubly stochastic if is has positive entries and both its rows and columns sum to 1.} The decomposition algorithm is basically the one presented above.\footnote{See \url{http://staff.science.uva.nl/~walton/Notes/Hall_Birkhoff.pdf} for a short presentation of the Birkhoff-von Neumann theorem and the decomposition algorithm.} 
\end{IEEEproof}

\begin{remark} In  the rest of the paper, we will refer to the algorithm in the proof of Lemma~\ref{rewriting} as the {\em decomposition algorithm}.   
\end{remark}

\begin{example}\label{matrixR}
Let $M=7$ and $\theta$ as in Example \ref{trivial}
and  suppose that $A_\tau\defeq a^3b^2c^2e^3f^4g^4h^2i$ is a  product in $\perm(\theta ^{\uparrow\matr{P}})$.  
  Then this product  corresponds to the following exponent matrix $R_\tau$ and the corresponding matrix $\theta^{R_\tau}\defeq (\theta_{ij}^{r_{ij}^\tau})$
  \begin{align}\label{Rtau}
&R_\tau\defeq\begin{bmatrix} 3&2&2\\0&3&4\\4&2&1\end{bmatrix},\quad \theta^{R_\tau}=\begin{bmatrix} a^3&b^2&c^2\\d^0&e^3&f^4\\g^4&h^2&i^1\end{bmatrix}.   \end{align}
 
Following the algorithm we obtain
\begin{align*} R_\tau=\begin{bmatrix} {\bf 3}&2&2\\0&{\bf 3}&4\\4&2&\fbox{\bf 1}\end{bmatrix}&\rightarrow  (aei)\rightarrow \begin{bmatrix} 2&\fbox{\bf 2}&2\\0&2&{\bf 4}\\{\bf 4}&2&0\end{bmatrix}\rightarrow (bfg)^2\rightarrow
\begin{bmatrix} 2&0&\fbox{\bf 2}\\0&{\bf 2}&2\\{\bf 2}&2&0\end{bmatrix} \rightarrow  (ceg)^2\rightarrow  \begin{bmatrix} \fbox{\bf 2}&0&0\\0&0&{\bf 2}\\0&{\bf 2}&0\end{bmatrix}\rightarrow (afh)^2. \end{align*}
So $a^3b^2c^2e^3f^4g^4h^2i =(aei)(bfg)^2(ceg)^2(afh)^2.$ 
\end{example} 
It can be easily checked that the  decomposition in Example \ref{matrixR} is unique. However, this is not always the case.  
Next we show an example where there are 3 possible decompositions. 
\begin{example} \label{matrixR2} Let $M=7$ and $\theta$ as in Example \ref{trivial}. 

Suppose that $A_\kappa\defeq a^3b^2c^2e^2f^3g^2g^2h^2i^3$ is a  permanent-product in $\perm(\theta ^{\uparrow\matr{P}})$ that 
corresponds to the following exponent matrix $R_\kappa$ and the corresponding matrix $\theta^{R_\kappa}\defeq (\theta_{ij}^{r_{ij}^\kappa})$
  \begin{align}\label{Rkappa}
&R_\kappa\defeq\begin{bmatrix} 3&2&2\\2&3&2\\2&2&3\end{bmatrix},\quad \theta^{R_\kappa}=\begin{bmatrix} a^3&b^2&c^2\\d^2&e^3&f^2\\g^2&h^2&i^3\end{bmatrix}.   \end{align}
 
Following the algorithm we obtain
\begin{align*} R_\kappa=\begin{bmatrix} {\bf 3}&2&2\\2&{\bf 3}&2\\2&2&\fbox{\bf 3}\end{bmatrix}&\rightarrow  (aei)^3\rightarrow \begin{bmatrix} 0&\fbox{\bf 2}&2\\2&0&{\bf 2}\\{\bf 2}&2&0\end{bmatrix}\rightarrow (bfg)^2\rightarrow
\begin{bmatrix} 0&0&\fbox{\bf 2}\\{\bf 2} &0& 0\\0& {\bf 2}&0\end{bmatrix} \rightarrow  (cdh)^2.
\end{align*}
So $a^3b^2c^2e^3f^4g^4h^2i =(aei)^3(bfg)^2(cdh)^2.$ 

However, we can also group the entries in the following way: 
\begin{align*} R_\kappa=\begin{bmatrix} {\bf 3}&2&2\\2&3&{\bf 2}\\2& \fbox{\bf 2}& 3\end{bmatrix}&\rightarrow  (afh)^2\rightarrow 
\begin{bmatrix} \fbox{\bf 1}& 2&2\\2&{\bf 3}&0\\ 2&0&{\bf 3}\end{bmatrix}\rightarrow (aei)\rightarrow
\begin{bmatrix} 0&2&\fbox{\bf 2}\\2& {\bf 2} & 0\\{\bf 2}& 0&2\end{bmatrix}  \rightarrow  (ceg)^2\rightarrow  \begin{bmatrix} 0&\fbox{\bf 2}&0\\{\bf 2} &0&0\\0&0&{\bf 2}\end{bmatrix}\rightarrow (bdi)^2. \end{align*}
So $a^3b^2c^2e^3f^4g^4h^2i =(afh)^2(aei)(ceg)^2(bdi)^2.$ 
Similarly, we can also group them in the following way: 
\begin{align*} R_\kappa=\begin{bmatrix} {\bf 3}&2&2\\2&3&{\bf 2}\\2& \fbox{\bf 2}& 3\end{bmatrix}&\rightarrow  (afh)\rightarrow 
\begin{bmatrix} \fbox{\bf 2}& 2&2\\2&{\bf 3}&1\\ 2&1&{\bf 3}\end{bmatrix}\rightarrow (aei)^2\rightarrow
\begin{bmatrix} 0&2&\fbox{\bf 2}\\2& {\bf 1} & 1\\{\bf 2}& 1&1\end{bmatrix}  \\&\rightarrow  (ceg)\rightarrow  \begin{bmatrix} 0&\fbox{\bf 2}&1\\{\bf 2} &0&1\\1&1&{\bf 1}\end{bmatrix}\rightarrow (bdi)\rightarrow\begin{bmatrix} 0&\fbox{\bf 1}&1\\1&0&{\bf 1}\\{\bf 1}&1&0\end{bmatrix}\rightarrow (bfg)\begin{bmatrix} 0&0&\fbox{\bf 1}\\{\bf 1}&0&0\\0&{\bf 1}&0\end{bmatrix}\rightarrow (cdh).\end{align*}
So $a^3b^2c^2e^3f^4g^4h^2i =(afh)(aei)^2(ceg)(bdi)(bfg)(cdh).$ 
It can be easily seen that these three decompositions are the only possible ones. 
\end{example} 
Therefore, the decomposition is not always unique, i.e., there are exponent matrices  for which the decomposition is unique and there are matrices  for which the decomposition is not unique. 
We will refer to a decomposition of an exponent matrix obtained by the decomposition 
algorithm as a {\em standard decomposition of the exponent matrix}. Similarly, we will refer to a decomposition of a product $\prod\limits_{j=1}^{m}\prod\limits_{l=1}^{m} (\theta_{jl})^{r_{jl}}$ into some product 
$\prod\limits_{\sigma \in {\mathcal S}_{m}} \left(\theta_{1\sigma(1)} \theta_{2\sigma(2)}  \cdots \theta_{m\sigma(m)}\right)^{t_{\tau\sigma}} $
as a {\em standard decomposition of the permanent-product}, as it corresponds to a standard decomposition of the exponent matrix. 

\subsection{Same-index decomposition of  a permanent-product} \label{sec:grouping}
The algorithm presented in the proof of Lemma~\ref{rewriting} provides a way to decompose the product $\prod\limits_{j=1}^{m}\prod\limits_{l=1}^{m} (\theta_{jl})^{r_{jl}}$ into a new product  
$\prod\limits_{\sigma \in {\mathcal S}_{m}} \left(\theta_{1\sigma(1)} \theta_{2\sigma(2)}  \cdots \theta_{m\sigma(m)}\right)^{t_{\tau\sigma}} $ 
but does not tell us exactly how to combine the entries $(\theta^{\uparrow\matr{P}})_{i\tau(i)}$ to obtain this decomposition.  Is there a way in which we can algorithmically  combine the indices of the sets $\alpha_{jl}^\tau$ to form the products $ \left(\theta_{1\sigma(1)} \theta_{2\sigma(2)}  \cdots \theta_{m\sigma(m)}\right)^{t_{\tau\sigma}} $ for all $\sigma \in {\mathcal S}_{m}$?  The answer is yes, as we explain in the next example of a concrete $\matr{P}$-lifting of $\theta$ from Example~\ref{matrixR} with $\matr{P}$ reduced.

Before presenting it, let us introduce a new matrix  $\overline{\alpha}_\tau \defeq (\overline{\alpha}_{jl}^\tau) $  obtained from $\alpha_\tau$ by substituting each index $(j-1)M+k$ in an entry set by  $k$, $k\in [M]$. Then the properties \eqref{alpha-r} of the matrix $\alpha_\tau$ translate into the following properties of the matrix $\overline{\alpha}_\tau$:
\begin{align}\label{alpha-new-r} 
&\overline{\alpha}_{jl}^\tau \bigcap  \overline{\alpha}_{jl'}^\tau =\emptyset,\; \forall j, l,l'  \in [m], l\neq l',   \quad   \bigcup_{l=1}^m  \overline{\alpha}_{jl}^\tau =[M].  
 \end{align}

The following example uses the matrix $\overline{\alpha}$ and provides a unique method of combining the indices $\overline{\alpha}_{jl}^\tau$ to obtain the desired decomposition of the product $A_\tau$. This method follows  the steps of  the algorithm that we described in Example~\ref{matrixR} for modifying the matrix $R_\tau$. 

 \begin{example}  \label{example-lift-coefficient} Let $\theta$ be the $3\times  3 $ matrix in Example~\ref{trivial}, $\matr{P}=(P_{ij}) \in  \mathcal{\overline P}_3^{3}$,  ${\theta}^{\uparrow\matr{P}}$ and $A_\tau=a^2bdf^2h^2i$ as follows: 
\begin{align*} 
 \matr{P}=(P_{ij}) \defeq \begin {bmatrix} I_3&I_3&I_3\\I_3&Q&Q^2
\\ I_3&I_3&Q^2\end {bmatrix}, \quad & Q\defeq \begin{bmatrix} 0&0&1\\1&0&0\\0&1&0\end{bmatrix},  \quad Q^2\defeq \begin{bmatrix} 0&1&0\\0&0&1\\1&0&0\end{bmatrix},\end{align*}  
\begin{align}{\theta}^{\uparrow\matr{P}}\defeq 
 \setlength{\arraycolsep}{3.2pt}\left[\begin {array}{ccc|ccc|ccc} \fbox{$a$}&0&0  &b&0&0  &c&0&0\\ 0&a&0& 0&\fbox{$b$}&0  &0&c&0\\0&0&\fbox{$a$}&0&0&b&0&0&c\\\hline
  d&0&0& 0&0&e &0&\fbox{$f$}&0\\ 0&\fbox{$d$}&0& e& 0&0 &0&0&f\\ 0&0&d& 0&e&0 &\fbox{$f$}&0&0
\\ \hline g&0&0& \fbox{$h$}&0&0 &0&i&0\\ 0&g&0&0&h&0&0&0&\fbox{$i$}\\ 0&0&g&0&0&\fbox{$h$}&i&0&0
\end {array}\right] =   &
\setlength{\arraycolsep}{3.4pt}\left[ \begin {array} {ccc|ccc|ccc}\fbox{$a_1$}&0&0  &b_1&0&0  &c_1&0&0\\ 0&a_2&0& 0&\circled{$b_2$}&0  &0&c_2&0\\0&0&\shade{a_3}&0&0&b_3&0&0&c_3\\\hline
  d_1&0&0& 0&0&e_1 &0&\fbox{$f_1$}&0\\ 0&\circled{$d_2$}&0& e_2& 0&0 &0&0&f_2\\ 0&0&d_3& 0&e_3&0 &\shade{f_3}&0&0
\\ \hline g_1&0&0& \fbox{$h_1$}&0&0 &0&i_1&0\\ 0&g_2&0&0&h_2&0&0&0&\circled{$i_2$}\\ 0&0&g_3&0&0&\shade{h_3}&i_3&0&0
\end {array}\right],  \label{example-lift} 
\end{align}
where $I_3$ denotes the identity matrix of size $3$ and  
the entries boxed in \eqref{example-lift} (left matrix)  correspond to the permutation $\tau$ that gives the product $A_\tau=a^2bdf^2h^2i$. In  \eqref{example-lift} (right matrix), we wrote the matrix  ${\theta}^{\uparrow\matr{P}}$ with its entries indexed by their row,  e.g., $a_1=a_2=a_3=a$  and $a_i$ is on the $i$th row of the first block $P_{11}$. 

The matrices $\alpha_\tau$, $\overline{\alpha}_\tau$ 
and $R_\tau$ are 
\begin{align*} \alpha_\tau&=\setlength{\arraycolsep}{4pt}\left[ \begin{array}{c|c|c}\{1,3\}&\{2\}&\emptyset\\\hline\{5\}&\emptyset&\{4,6\}\\\hline\emptyset&\{7,9\}&\{8\} \end{array}\right],  \quad  
\overline{\alpha}_\tau =\setlength{\arraycolsep}{4pt}\left[\begin{array}{c|c|c} \fbox{1}~\shade{3}&\circled{$2$}&\emptyset\\\hline\circled{$2$}&\emptyset&\fbox{1}~\shade{3}\\\hline\emptyset&\fbox{1}~\shade{3}&\circled{$2$} \end{array}\right], \quad R_\tau= \begin {bmatrix} 2&1&0\\1&0&2\\
0&2&1\end {bmatrix},\end{align*}
where, for simplicity in writing, we omit the set parentheses in $\overline{\alpha}_\tau$. 
Note that $\overline{\alpha}_\tau$ corresponds to the row indices of the boxed entries in \eqref{example-lift} (left matrix) that are illustrated through indexed entries in \eqref{example-lift} (right matrix). 
In the matrix $ \overline{\alpha}_\tau$, we use circles, boxes  and shaded boxes to show how to group the entries of \eqref{example-lift} (left matrix) that appear in $A_\tau$, as follows.  We group together entries in $ {\theta}^{\uparrow\matr{P}}$ in rows  indexed by the  circled entries in $ \overline{\alpha}_\tau$, and we group together entries in $ {\theta}^{\uparrow\matr{P}}$ in rows  indexed by the boxed entries  in $ \overline{\alpha}_\tau$, thus obtaining  a unique rewriting of the product $A_\tau$ as    $A_\tau =(afh)^2(bdi)$, in correspondence to the rewriting steps of  matrix $R_\tau$. 
In terms of the indexed entries of ${\theta}^{\uparrow\matr{P}}$, the above grouping corresponds to
$A_\tau = (a_1f_1h_1)(a_3f_3h_3)(b_2d_2i_2)$ which is exemplified through circles, boxes and shaded boxes in the version of ${\theta}^{\uparrow\matr{P}}$ with indexed entries in~\eqref{example-lift} (right matrix).   \end{example}

Is a  decomposition like the one drawn in $\overline{\alpha}_\tau$ of  Example~\ref{example-lift-coefficient} always possible? The answer is yes due to the following simple fact. Each row and column of  ${\theta}^{\uparrow\matr{P}}$ participates with exactly one element to a permanent-product. In the matrix  ${\theta}^{\uparrow\matr{P}}$ of~\eqref{example-lift}, once we choose $d$ on the second column, or, equivalently, $d_2$, none of the entries $a_2$ or $g_2$ on that column can be part of the permanent-product anymore and,  therefore, the second row of matrix $P_{11}$  (where $a_2$ is positioned) and the second row of the matrix $P_{31}$ (where $g_2$ is positioned) must contribute each with exactly one entry other than the entries $a_2$ and $g_2$ that are not allowed. These are the boxed entries $b_2$ and $i_2$. We group these entries with $d_2$ uniquely and continue the same way to group 
each of the $a$ entries with the entries $f$ and $h$ that are on the two rows associated with the other two entries on the columns of the entries $a$ to obtain  $(a_1f_1h_1)$ and $(a_3f_3h_3)$. 

In terms of the entries of the matrix $\overline{\alpha}_\tau$, this corresponds to the grouping we showed in Example~\ref{example-lift-coefficient} because the matrix $\matr{P}$ is reduced, so the first matrices $P_{l1}$ in each row and $P_{1l}$ in each column are equal to the the identity matrix, for all $l\in [m]$.  Therefore, for each of the first $M$ columns, the nonzero entries on the $j$th column are all positioned on the $j$th row of the matrices  $P_{l1}$,  for all  $l\in [m]$. 
Of course, this is not valid for a column that is not among the first $M$. Indeed, the boxed $i$ of ${\theta}^{\uparrow\matr{P}}$ in \eqref{example-lift} is on row 2 of matrix $P_{33}$ and  has the nonzero entries on rows 3 of matrix $P_{13}$ and 2 of matrix  $P_{23}$. However, it still holds that the rows corresponding to these non-zero entries must contribute to the product with one entry exactly that cannot be on the column of $i$. In this case,  $d_2$ on position $(2,2)$ in $P_{21}$ and $a_3$ on position $(3,3)$ of $P_{11}$ are these entries. We can group these together as well. In fact any such grouping of three where two of them are on the rows corresponding to the non-chosen entries of the column of the third of the group is a good association;  the permanent-product $A_\tau$ is then a product  of some of these three-products with the property that the entries in the products are taken only once and they cover all the entries in the permanent-product $A_\tau$ (i.e., they form a partition). 
Such a partition is surely given by the three-sets of the boxed entries in the first $M$ columns, because each of these sets must be disjoint and they are exactly $M$, the number of boxed entries from the first $M$ columns, so the union of all entries in these products is equal to all entries in the product $A_\tau$. In fact, any three-sets  associated to the boxed entries in a set  $(j-1)M+1, \ldots, jM$  of columns corresponds to a partition of the entries in $A_\tau$. For simplicity, however, we choose the partition corresponding to the first $M$ columns, or, equivalently, to the matrix $\overline{\alpha}_\tau$. We call this decomposition {\em same-index decomposition}. 

Therefore, the {\em same-index decomposition of a permanent-product}  in ${\theta}^{\uparrow\matr{P}}$ is the writing  of the permanent-product as a product of $M$ sub-products of $m$ entries in $\theta$ each  indexed by the same row index, e.g., $(a_1f_1h_1)(b_2d_2i_2)( a_3f_3h_3)$.

\subsection{The relation between the exponent matrix decomposition and the permanent-product same-index decomposition} \label{sec:index:grouping}
In this section we will revisit the setting of  Example  \ref{matrixR2} in order  to understand how the grouping described in Section \ref{sec:grouping} determines the type of the decomposition into $M$ products of permanent-products of $\theta$ in the decomposition algorithm of the exponent matrix described in Section \ref{sec:rewriting}. 
\begin{example} \label{index-grouping} Let $M=7$ and let $\theta$ and $A_\kappa\defeq a^3b^2c^2e^2f^3g^2g^2h^2i^3$ as in Example \ref{matrixR2}. We saw that there were three possible decompositions of   $A_\kappa$ in permanent-products 
as follows 
$$A_\kappa= (afh)^2(aei)(ceg)^2(bdi)^2 =(afh)(aei)^2(ceg)(bdi)(bfg)(cdh) =(aei)^3(bfg)^2(cdh)^2.$$
How are these three possible  decompositions of the exponent matrix visible in the same-index decomposition of a given permanent-product described in Section \ref{sec:grouping}? We can assume for simplicity (and without loss of generality) that the exponent $r_{11}=3$ corresponds to the row indices $\{1,2,3\}$ of the entries in $P_{11}$ that appear in $A_\kappa$. We have three possible scenarios for how these row indices can be  combined with the indices of the entries in 
$\begin{bmatrix} P_{22}&P_{23} \\P_{32}&P_{33}\end{bmatrix}$ with associated exponent matrix $\begin{bmatrix} 3&2\\2&3\end{bmatrix} $ (modulo some permutations of indices)
such that the overall exponent matrix is $R_\kappa$:
\begin{align*}  
\overline{\alpha}_\kappa =\setlength{\arraycolsep}{4pt}\left[\begin{array}{c|c|c} \fbox{1}~\circled{$2$}~\shade{3}&& \\ \hline &\fbox{1}~\circled{$2$}~\shade{3}&\\\hline &&\fbox{1}~\circled{$2$}~\shade{3} \end{array}\right];\setlength{\arraycolsep}{4pt}\left[\begin{array}{c|c|c} \fbox{1}~\circled{$2$}~\shade{3}&& \\ \hline &\fbox{1}~\circled{$2$}&\shade{3}\\\hline &\shade{3}&\fbox{1}~\circled{$2$} \end{array}\right];
\setlength{\arraycolsep}{4pt}\left[\begin{array}{c|c|c} \fbox{1}~\circled{$2$}~\shade{3}&& \\ \hline &\fbox{1}&\circled{$2$}~\shade{3}\\\hline &\circled{$2$}~\shade{3} &\fbox{1}\end{array}\right].
\end{align*}
These  correspond to the following exponent matrices: 
\begin{align*}&R_{\kappa, a^3e^3i^3} =\begin{bmatrix} 3&&\\&3&\\&&3\end{bmatrix}; \quad R_{\kappa, a^3e^2fhi^2} =\begin{bmatrix} 3&&\\&2&1\\&1&2\end{bmatrix};  \quad R_{\kappa, a^3ef^2h^2i} =\begin{bmatrix} 3&&\\&1&2\\&2&1\end{bmatrix}. \end{align*}
The remaining indices are  uniquely determined in the way shown in Example \ref{matrixR2} so we omit them from the matrices above.  

Equivalently,  we have the following possible same-index decompositions: 
 \begin{align*} 
 a_1a_2a_3~ e_1e_2e_3~i_1i_2i_3&= (a_1e_1i_1)(a_2e_2i_2)(a_3e_3i_3)\\
 a_1a_2a_3 ~e_1e_2f_3~i_1i_2h_3&= (a_1e_1i_1)(a_2e_2i_2)(a_3f_3h_3)\\
 a_1a_2a_3 ~e_1f_2f_3~i_1h_2h_3&= (a_1e_1i_1)(a_2f_2h_2)(a_3f_3h_3).
 \end{align*}  
Therefore, when fixing the indices of $P_{11}$ to $\{1,2,3\}$, there are  3  {\em non-equivalent} ways in which the exponent matrix $R_\kappa$ in \eqref{Rkappa} can occur, where by {\em non-equivalent} we mean that the matrices $ \overline{\alpha}_\kappa$ do not map into each other after applying some permutation on the set of row indices $[M]$. 

However, in the case of $A_\tau=a^3b^2c^2d^0e^3f^4g^4h^2i^1$ in Example \ref{matrixR} with the exponent $R_\tau$ given in \eqref{Rtau} 
 we can only have 
 \begin{align*}  
\overline{\alpha}_\kappa =\setlength{\arraycolsep}{4pt}\left[\begin{array}{c|c|c} \fbox{1}~\circled{$2$}~\shade{3}&& \\ \hline &\fbox{1}&\circled{$2$}~\shade{3}\\\hline &\circled{$2$}~\shade{3}&\fbox{1} \end{array}\right]
\end{align*}
(or equivalent  matrices) due to the entry of $1$ in the position $(3,3)$ of $R_\tau$ and $0$ in the position $(2,1)$. Indeed, as explained in Section \ref{sec:grouping},  if the  entry from  $P_{3,3}$ is,  for example,  on row $1$,  the entries on the row $8$ (the first of the second row of blocks) must contribute to the permanent-product with an entry from the matrices  $P_{2,1}$ or $P_{2,2}$. Since $r_{21}=0$,   it  implies that  
 the entry on the first row of $P_{2,2}$ is also in the permanent-product. Similarly, the entry on the first row of $P_{1,1}$ will also appear in the product, and thus we have the unique scenario (modulo permutations) presented above and the product 
 $$a_1a_2a_3~ e_1f_2f_3~h_2h_3i_1= (a_1e_1i_1)(a_2f_2h_2)(a_3f_3h_3). $$
 \end{example}
 So far, in all our examples the same-index decomposition of a permanent-product is equal to its standard decomposition. In the following section, we see that this is not always the case. 
 \subsection{Decompositions that contain illegal sub-products}\label{sec:illegal:products} 
 
Note that in Example \ref{index-grouping}  one of the following decompositions in $ \overline{\alpha}_\kappa$ associated with the entry $3$ in the position $(1,1)$ of $R_\kappa$  could also occur (and their equivalent version):
 
{\footnotesize\begin{align*}  
\setlength{\arraycolsep}{3pt}\left[\begin{array}{c|c|c} \fbox{1}~\circled{$2$}~\shade{3}&& \\ \hline &\fbox{1}~\circled{$2$}&\shade{3}\\\hline &\fbox{1}&\circled{$2$}~\shade{3} \end{array}\right],
\setlength{\arraycolsep}{3pt}\left[\begin{array}{c|c|c} \fbox{1}~\circled{$2$}~\shade{3}&& \\ \hline &\shade{3}~\circled{$2$}&\fbox{1}\\\hline &\shade{3}&\circled{$2$}~\fbox{1} \end{array}\right],
\setlength{\arraycolsep}{3pt}\left[\begin{array}{c|c|c} \fbox{1}~\circled{$2$}~\shade{3}&& \\ \hline &\fbox{1}&\circled{$2$}~\shade{3}\\\hline &\fbox{1}~\circled{$2$}&\shade{3} \end{array}\right],
\setlength{\arraycolsep}{3pt}\left[\begin{array}{c|c|c} \fbox{1}~\circled{$2$}~\shade{3}&& \\ \hline &\shade{3}&\circled{$2$}~\fbox{1}\\\hline &~\shade{3}\circled{$2$}&\fbox{1} \end{array}\right],
\end{align*}}
yielding the following  permanent-products of ${\theta}^{\uparrow\matr{P}}$: 
  \begin{align*} 
 a_1a_2a_3 ~e_1e_2f_3~h_1i_2i_3&= (a_1e_1h_1)^{\ill}(a_2e_2i_2)(a_3f_3i_3)^{\ill}= (a_1e_1i_3)(a_2e_2i_2)(a_3f_3h_1),\\
  a_1a_2a_3 ~f_1e_2e_3~i_1i_2h_3&= (a_1f_1i_1)^{\ill}(a_2e_2i_2)(a_3e_3h_3)^{\ill}= (a_1f_1h_3)(a_2e_2i_2)(a_3e_3i_1),\\
 a_1a_2a_3 ~e_1f_2f_3~h_1h_2i_3&= (a_1e_1h_1)^{\ill}(a_2f_2h_2)(a_3f_3i_3)^{\ill}= (a_1e_1i_3)(a_2f_2h_2)(a_3f_3h_1),\\
a_1a_2a_3 ~e_3f_2f_1~h_3h_2i_1&= (a_1f_1i_1)^{\ill}(a_2f_2h_2)(a_3e_3h_3)^{\ill}= (a_1f_1h_3)(a_2f_2h_2)(a_3e_3i_1). 
 \end{align*}  
In this case,  not all  of the  products  of $3$ entries of the same index correspond to permanent-products in the matrix $\theta$;   we marked with $\dagger$  the ones that do not,  for example, $(a_1   e_{1}  h_1)^{\ill}$ corresponds to $aeh$ in $\theta$ which is not a   permanent-product. We call such a product {\em illegal}.  This illegal three-product needs to be grouped with another illegal three-product in the same grouping, in this case $(a_3f_3i_3)^{\ill}$,   and rearranged as $(a_1e_1i_3)(a_3f_3h_1)$ to obtain  a standard  decomposition, i.e., a product of permanent-products of $\theta$. We call these sub-products that correspond to a permanent-product in $\theta$ {\em legal}.

\begin{example}  \label{matrix3by6}
Let $\theta$ the $3\times 3$ matrix from Example~\ref{trivial}, let $M=6$,  and
let
{\footnotesize\begin{align*}
 {\theta}^{\uparrow\matr{P}}\defeq \setlength{\arraycolsep}{3pt}\left[\begin{array}{cccccc|cccccc|cccccc}
\circled{$\bf a_1$}&0    &0    &0  &0  &0   &               b_1&0&0 &0&0&0&                            c_1&0&0&0&0&0\\ 
                     0&a_2&0    & 0 &0  &0   &                     0&b_2&0&0&0&0&               0&\circled{$c_2$}&0&0&0&0\\
                     0&0    &a_3&0  &0  &0   &           0&0&b_3&0&0& 0&                               0&0&\fbox{$c_3$}&0&0&0\\
                     0& 0   &0    &\fbox{$\bf a_4$}&0&0   &          0& 0&0&b_4&0&0&                               0&0&0&c_4&0&0\\
                     0&     0&0   &0&\shade{\bf a_5}&0   &           0&0&0&0&b_5&0&                               0&0&0&0&c_5&0\\
                     0&     0&0   &0&0    &a_6&            0&0&0&0&0&\shade{b_6}&                               0&0&0&0&0&c_6\\\hline
                 d_1&0     &0   &0&0    &0 &                   0&0&e_1 &0&0&0&                            \circled{$\bf f_1$}&0&0&0&0&0\\ 
                     0&\circled{$d_2$} &0   & 0&0   &0 &                     0&0&0&0&0&e_2&               0&f_2&0&0&0&0\\
                     0&0     &d_3&0&0   &0 &           0&0&0&0&\fbox{$e_3$}& 0&                               0&0&f_3&0&0&0\\
                     0&0     &0    &d_4&0 &0 &          0& 0&0&e_4&0&0&                               0&0&0&0&\fbox{$\bf f_4$}&0\\
                     0&0& 0&0&d_5&0&               e_5&0&0&0&0&0&                               0&0&0&\shade{\bf f_5}&0&0\\
                   0&0&0&0&0&d_6&                  0&\shade{e_6}&0&0&0&0&                               0&0&0&0&0&f_6\\\hline
                 g_1&0&0 &0&0&0 &                  0&0& \circled{$\bf h_1$}&0&0 &0&                i_1&0&0&0&0&0\\ 
                  0&g_2&0& 0&0&0 &                 \circled{$h_2$}&0&0&0&0&0&             0&i_2&0&0&0&0\\
                  0&0&\fbox{$g_3$}&0&0&0&            0&0&0&0&h_3& 0&                               0&0&0&0&i_3&0\\
                  0&0&0&g_4&0&0&          0& 0&0&0&0&h_4&                               0&0&0&0&0&\fbox{$\bf i_4$}\\
                 0&0& 0&0&g_5&0&           0&0&0&\shade{\bf h_5}&0& 0&                              0&0&i_5&0&0&0\\
                   0&0&0&0&0&\shade{g_6}&            0&h_6&0&0&0&0&                               0&0&0&i_6&0&0
 \end{array}\right],\end{align*}}%
with the entries of the product $A_\tau\defeq a_1a_{4}a_{5}b_{6}c_2c_{3}d_2e_{3}e_{6}f_1f_{4}f_{5}g_{3}g_{6}h_1h_2h_{5}i_{4}= a^3bc^2de^2f^3g^2h^3i$
highlighted in the matrix ${\theta}^{\uparrow\matr{P}}$. All entries in a block $P_{ij}$ are equal to the entry $\theta_{ij}$, for example, $a_1=a_2=a_{3}=a_{4}=a_{5}=a_6=a,$ etc. We use the index $l$ for an entry in $P_{ij}$ to denote the row position of that entry in $P_{ij}$, for example, $a_{4}$ is on the $4$th row of $P_{11}$. The indices for the entries are helpful when describing $\tau$.  We use boxes, shades, circles and bold faced with circles,  boxes, and shades, respectively,  to draw the entries of $A_\tau$  so that the decomposition according to the matrix $ \overline{\alpha}_\tau$, i.e.,  the  same-index decomposition, is visible. This means that all entries in the permanent-product  $A_\tau$ of the same index will have the same shape/color.   The following matrices can be computed. 
 \begin{align*}  R_\tau =\begin{bmatrix} 3&1&2\\ 1&2&3\\ 2&3&1\end {bmatrix}\,;\quad \overline{\alpha}_\tau =
 \left[\begin{array}{c|c|c} \circled{$\bf 1$}~ \boxed{\bf 4}~ \shade{\bf 5} &\shade{6}&\circled{$2$}~ \boxed{3} \\\hline\circled{$2$}&\boxed{3}~\shade{6}&\circled{$\bf 1$}~\boxed{\bf 4}~\shade{\bf 5}\\\hline \boxed{3}~ \shade{6}&\circled{$\bf 1$}~\circled{$2$}~\shade{\bf 5}&\boxed{\bf 4} \end{array}\right].  
\end{align*}

The same-index decomposition
corresponding to the grouping of the matrix $\overline{\alpha}_\tau$  is
\begin{align*} 
 &
 A_\tau=(a_1   f_1   h_1) 
 (d_2   c_2   h_2)
  (g_{3}  c_{3}  e_{3}) 
  (a_{4} { f_{4}}~  i_{4})^\ill 
  (a_{5}  f_{5}  h_{5})
  (g_{6}  b_{6} {e_{6}}~)^\ill.
\end{align*}
 Here we have an example in which  not all  products  of $3$ entries of the same index in a permanent-product of ${\theta}^{\uparrow\matr{P}}$ are legal, i.e., they correspond to permanent-products in the matrix $\theta$.  As before,  we marked with $\ill$  the illegal ones,  for example, $(a_4   f_{4}  i_4)^\ill$ corresponds to $afi$, which is not a permanent-product in $\theta$. This three-product needs to be grouped with another illegal three-product in the same decomposition  and rearranged as follows. 
\begin{align*} &(a_4   f_{4}  i_4 )^\ill(g_6 b_6 e_6)^\ill
 = (a_4e_6i_4)(g_6b_6f_4)= (aei)(gbf). 
 \end{align*} 
 This results in the following standard decomposition of the permanent-products $A_\tau$  that is not equal to its same-index decomposition, i.e., into a product that contains only legal terms, although some of them contain sub-products with
 indices that are not all the same:
  $$A_\tau=(a_1   f_1   h_1) 
 (d_2   c_2   h_2)
  (g_{3}  c_{3}  e_{3}) 
  (a_{4} { e_{6}} i_{4}) 
  (a_{5}  f_{5}  h_{5})
  (g_{6}  b_{6} {f_{4}}).$$
  \vspace*{-0.4cm}
  \par
    \end{example}

\subsection{Mapping illegal products into legal products}\label{sec:mapping} 

In this section we will show that we can always assume that all permanent-products in $\theta^{\uparrow\matr{P}}$ are products of $\theta$-permanent-products by showing that any permanent-product of ${\theta}^{\uparrow\matr{P}}$ containing some illegal sub-products can  be mapped uniquely into some product of $M$ same-index permanent-products of $\theta$. In addition, this product has the same exponent matrix as the original  permanent-product but is not a permanent-product of ${\theta}^{\uparrow\matr{P}}$. This way, we establish
a one-to-one correspondence between permanent-products of $\theta^{\uparrow\matr{P}}$ and products of $M$ permanent-products in $\theta$. 

We revisit Example \ref{matrix3by6} to exemplify this correspondence. 
 \begin{example} \label{mapping1} Let $\theta$, $\theta^{\uparrow\matr{P}}$ and $A_\tau$ be like  in Example~\ref{matrix3by6}. 
Recall that the same-index decomposition of $A_\tau$ in Example~\ref{matrix3by6} was $A_\tau=(a_1   f_1   h_1) (d_2   c_2   h_2) (g_{3}  c_{3}  e_{3})  (a_{5}  f_{5}  h_{5})(a_{4}  f_{4}  i_{4})^\ill (g_{6}  b_{6}  e_{6})^\ill 
$, which contained two illegal sub-products $(a_{4}  f_{4}  i_{4})^\ill$ and $ (g_{6}  b_{6}  e_{6})^\ill$ that were combined  to obtain $(a_{4} { e_{6}} i_{4}) (g_{6}  b_{6} {f_{4}})$. Note that this combination is unique; no other combination resulting in legal sub-products, i.e., in permanent-products in $\theta$, is possible between the two products.  Each $\theta$-permanent-products $(a_{4} { e_{6}} i_{4})$ and $(g_{6}  b_{6} {f_{4}})$ contains the combined indices $4$ and $6$.   

Let  $A'_\tau \defeq (a_1   f_1   h_1) (d_2   c_2   h_2) (g_{3}  c_{3}  e_{3})  (a_{5}  f_{5}  h_{5}) (a_{4}  e_{4}  i_{4})(g_{6}  b_{6}  f_{6}) $ be the (unique) product of $M$ same-index $\theta$-permanent-products  starting with $a_1,d_2,g_3,a_4, a_5$ and $ g_6$ with the same  exponent matrix $R_\tau$.    Map $A_\tau\mapsto A'_\tau.$
 We observe that $A'_\tau$ cannot be a permanent-product if $A_\tau$ is.  Indeed, the two products are equal in all but 2 positions, therefore, if the two were both permanent-products, then $\theta^{\uparrow\matr{P}}$ would need to have a  $2\times 2$ submatrix $\begin{bmatrix} e_4&f_4\\e_6&f_6\end{bmatrix},$ which is not allowed as no two $e$ entries (and no two $f$ entries) are on the same row or column (they are entries in $eP_{22}$, respectively,  $fP_{23}$, where $P_{22}$ and $P_{23}$ are permutation matrices).
Hence the correspondence $A_\tau\mapsto A'_\tau$ is an instance of the desired correspondence between the permanent-products in $\theta^{\uparrow\matr{P}}$ that have illegal sub-products in their  same-index decompositions, and products of $M$ permanent-products in $\theta$ that are not permanent-products in $\theta^{\uparrow\matr{P}}$. In addition, since a permanent-product in $\theta^{\uparrow\matr{P}}$ that does not contain any illegal sub-products has its same-index decomposition equal to its standard decomposition, it can be mapped trivially into itself. This way, we obtain a map from the set of all permanent-products in  $\theta^{\uparrow\matr{P}}$ into the set of all $M$ products of permanent-products in $\theta$.  
\end{example} 

This correspondence illustrated in the previous example can be generalized to all permanent-products of ${\theta}^{\uparrow\matr{P}}$ with same-index decompositions that contain some illegal sub-products in the following way. 
\begin{itemize}
\item Let $\theta$ be an $m\times m$ non-negative matrix and ${\theta}^{\uparrow\matr{P}}$ be a reduced matrix of degree $M$. 
\item Let $\tau$ be a permutation on $[mM]$ and $A_\tau$ be a permanent-product in ${\theta}^{\uparrow\matr{P}}$ that is not trivially zero.  Let $R_\tau$ be its exponent matrix. 
\item
Write $ A_\tau$ as the same-index  decomposition;  $A_\tau$ can or not contain illegal same-index sub-products, i.e., products of $m$ entries in $\theta$  of the same index that are not permanent-products in $\theta$. 
\item
List all distinct  products of $M$ {\em same-index permanent-products} in $\theta$  corresponding to  all  standard decompositions of  $R_\tau$ that start with the entries in  $A_\tau$ that are in the first $M$ columns of ${\theta}^{\uparrow\matr{P}}$. Call them  $A'_{\tau, 1}, \ldots, A'_{\tau, l}$ and reorder,  if needed, the entries in the sub-products of  $A_\tau$ and $A'_{\tau, 1}, \ldots, A'_{\tau, l}$  such that the entries from the first $M$ columns are always first in the subproduct, followed by the entries ordered  by the row index in $\theta$ increasingly from $1$ to $m$  and such that the indices  of the $\theta$-permanent-products are ordered increasingly from $1$ to $M$. 
\end{itemize}
This procedure, henceforth called {\em standard mapping}, is formalized in the following lemma.
Several examples can be found in Appendix \ref{sec:app:map}.

\begin{lemma}[Standard mapping]\label{big-lemma} 
Initially, set ${\mathcal L} :=\{A'_{\tau, 1}, \ldots, A'_{\tau, l}\}$. 

\fbox{Start} Let $0\leq s\leq M$ and $1\leq t<m$ be such that 
\begin{itemize}
\item $A_\tau$ and each $A'_{\tau, j}\in {\mathcal L}$  have  their  first $s$ $\theta$-permanent-products equal and 
\item $A_\tau$ and  each $A'_{\tau, j}\in  {\mathcal L}$  have their $(s+1)$th $\theta$-permanent-products either equal in the first $t$ entries or have all of the first $t$ entries distinct  except  for the first entry and 
\item $A_\tau$ and $A'_{\tau, i} \in {\mathcal L}$ have their $(s+1)$th $\theta$-permanent-product equal in the $(t+1)$th entry, while there exists  $A'_{\tau, j} \neq A'_{\tau, i}$,   such that $A_\tau$ and  $A'_{\tau, j}$ have the $(s+1)$th $\theta$-permanent-product distinct in the $(t+1)$th entry. 
\end{itemize} 
Let    $ \{A'_{\tau, j_1}, \ldots , A'_{\tau, j_k}\}\subset \{A'_{\tau, 1}, \ldots, A'_{\tau, l}\}, 1\leq k<l$, such that  $A_\tau$ and each $A'_{\tau, j_n}$, $n\in [k]$,  have their $(s+1)$th $\theta$-permanent-product equal in the $(t+1)$th entry. 

Map  $A_\tau\mapsto A'_{\tau, i}$ if $k=1$, otherwise update ${\mathcal L}:={\mathcal L}_k$ and repeat the steps from \fbox{Start}.

Then, this map is a well-defined one-to-one (injective) map from the set of all permanent products of ${\theta}^{\uparrow\matr{P}}$ of a certain exponent matrix to the set of all products of $M$ $\theta$-permanent-products of the same exponent matrix.  This gives a one-to-one map from the set of all permanent-products in  ${\theta}^{\uparrow\matr{P}}$ to the set of all products of $M$ $\theta$-permanent-products.  
\end{lemma} 
\begin{IEEEproof} The fact that the map is well-defined is easy to see since there can only be one matrix $A'_{\tau, i} $ satisfying the conditions, while the existence of this matrix is ensured by the decomposition algorithm presented in Section \ref{sec:rewriting}. Indeed, the exponent matrix decomposing algorithm guarantees the existence of the list of products of $\theta$-permanent-product, which has its cardinality at least one, and at the same time, guarantees the existence of  a standard decomposition of the permanent-product  into legal sub-products not necessarily of the same index obtained from its same-index decomposition; this can be mapped into a  product of  same-index $\theta$-permanent-products, thus guaranteeing the existence. 
The fact that no two permanent-products can be mapped into the same $A'_{\tau, i}$ is also ensured by the conditions of the mapping; if two different permanent products $A_\tau$ and $A_\nu$ map into the same $A'_{\tau, i}$, then they must have a first entry in which they differ; this entry must be  necessarily after the first $s$ entries. This means, however, that  there must exist an $A'_{\tau, j}$ that shares with $A_\nu$ that entry but not with  $A'_{\tau, i}$. Therefore, $A_\nu$ cannot get mapped into the same $A'_{\tau, i}$ as $A_\tau$, proving that the function is one-to-one. In addition, if $A_\tau$ contains illegal same-index sub-products, then $A'_{\tau, i}$ such that $A_\tau\mapsto A'_{\tau, i}$ cannot be a permanent-product in ${\theta}^{\uparrow\matr{P}}$. To see this,
erase from ${\theta}^{\uparrow\matr{P}}$ all rows and columns corresponding to the entries that the two share. Suppose that there are $k$ entries in which the two products are different, say, $x_1, x_2, \ldots, x_k$ in $A_\tau$ and $x'_1, x'_2, \ldots, x'_k$ in $A'_{\tau,i}. $ Because the two products $A_\tau, A'_{\tau, i}$ have the same exponent  matrix, so do the two products $x_1x_2\ldots x_k$ and $x'_1 x'_2 \ldots x'_k$. Therefore, in each block in which there exists some  $x_i$, $i\in [k]$, there must exist  also a $j\in [k]$ such that $x'_j$ is also in that block. We can reorder $x'_1, x'_2, \ldots, x'_k$  so that each $x'_l$ is in the same block as $x'_l$. Note that there can be more entries in one block, but to each entry $x_l$ corresponds a unique entry $x'_l$ in the same block. Since there is only one column in the $k\times k$ submatrix crossing the term $x_i$ and since  $x'_j\not \in\{x_1,\ldots x_k\} $, we obtain that $x_i$ and $x'_j$ must be on the same column which contradicts the fact that  the block is  a weighted  permutation matrix.

Therefore, if $A_\tau$ contains illegal same-index sub-products, then it is mapped through the above mapping into a product  $A'_{\tau, i}$ that is not a permanent-product in ${\theta}^{\uparrow\matr{P}}$. This also implies that an  all-legal permanent-product $A_\tau$   and a permanent-product in 
containing some illegal same-index sub-products $A_\kappa$ do not map into the same product of $M$  $\theta$-permanent-products, which in this case would be $A_\tau$. Indeed,  if $A_\tau$ does not contain any illegal sub-products, i.e., it is a product of $M$ $\theta$-permanent-products, then $A_\tau=A'_{\tau, i}$, for some $i$,  and the mapping corresponds to $A_\tau\mapsto A_\tau$ as expected. 

Such a mapping can be defined for each exponent matrix, which proves the existence of the overall one-to-one map 
from the set of all permanent-products in  ${\theta}^{\uparrow\matr{P}}$ to the set of all products of $M$ $\theta$-permanent-products.
\end{IEEEproof}

\subsection{Upper bounding the permanent of a  lifting of a matrix} \label{sec:coefficient} 
The mapping in Section \ref{sec:mapping} allows us to compute,  for a fixed exponent matrix $R=(r_{ij})$, the coefficient of $\prod\limits_{j=1}^{m}\prod\limits_{l=1}^{m} (\theta_{jl})^{r_{jl}}$   in $\perm(\theta^{\uparrow\matr{P}})$,  or, equivalently, the maximum possible number of permutations $\tau \in {\mathcal S}_{mM}$ such that   $A_\tau= \prod\limits_{j=1}^{m}\prod\limits_{l=1}^{m} (\theta_{jl})^{r_{jl}}$ is a permanent-product with exponent matrix $R$ that is not trivially-zero, and, using this, to prove the upper bound $\perm(\theta^{\uparrow\matr{P}})\leq \perm(\theta)^M.$

The following corollary is an immediate consequence of the one-to-one mapping. 
\begin{corollary} \label{cor:coeff}
Let $R=(r_{ij})$ be an exponent matrix of some permanent-product in $\perm(\theta^{\uparrow\matr{P}})$. For each $\tau \in {\mathcal S}_{mM}$ with  $A_\tau= \prod\limits_{j=1}^{m}\prod\limits_{l=1}^{m} (\theta_{jl})^{r_{jl}}$,  let $A'_{\tau, 1}, \ldots, A'_{\tau, l}$ be the possible products of $M$ $\theta$-permanent-products associated with $R$.  For each $j\in [l]$, denote by $N_{\tau,j}$ the  number of  products of $M$ $\theta$-permanent-products that are equivalent to $A'_{\tau, j}$, i.e., they can be obtained from $A'_{\tau, j}$ by applying an $M$-permutation on the indices. Then, the coefficient of $\prod\limits_{j=1}^{m}\prod\limits_{l=1}^{m} (\theta_{jl})^{r_{jl}}$ in $\perm(\theta ^{\uparrow\matr{P}})$ is upper bounded by $\sum\limits_{j=1}^l N_{\tau,j}$.   
\end{corollary} 

The following lemma determines the number $N_{\tau,j}$ for all $j\in [l]$. 

\begin{lemma}\label{the-lemma} 
For each $j\in [l]$ and $\sigma\in {\mathcal S}_{m}$,   let $0\leq t_{j, \sigma}\leq M$ such that $\sum\limits_{\sigma \in {\mathcal S}_{m}} t_{j,\sigma}=M$ and 
$A'_{\tau,j}=
\prod\limits_{\sigma \in {\mathcal S}_{m}} \left(\theta_{1\sigma(1)} \theta_{2\sigma(2)}  \cdots \theta_{m\sigma(m)}\right)^{t_{j,\sigma}}.$  
Then $N_{\tau,j} ={M \choose {\vect{t_{j}}}}$ where 
${M \choose {\vect{t_{j}}}}$ is the multinomial coefficient associated with the vector $\vect{t_{j}}\defeq (t_{j,\sigma})_{\sigma \in {\mathcal S}_{m}}.$
  \end{lemma} 
    \begin{IEEEproof} 
     The entries that lie in the first $M$ columns of  $\theta^{\uparrow\matr{P}}$ uniquely determine the way the products of  $\theta$-permanent-products   $\left(\theta_{1\sigma(1)} \theta_{2\sigma(2)}  \cdots \theta_{m\sigma(m)}\right)^{t_{j,\sigma}}$ are formed. We can choose these in ${M \choose {\vect{t_{j}}}}$ ways.
        \end{IEEEproof}

 The main result of the paper now follows immediately.  
  \begin{theorem} \label{the-theorem} Let $\theta=(\theta_{ij})$ be a non-negative matrix of size $m\times m$ and let $\matr{P}=(P_{ij})  \in   \mathcal{P}_M^{m}$. Then 
$$\perm(\theta ^{\uparrow\matr{P}})\leq  \perm(\theta)^M.
$$
\end{theorem} 
\begin{IEEEproof}
 The upper bound  follows immediately from Lemma~\ref{the-lemma}  and the  expansion of  $\perm(\theta)^M$ as 
\begin{align*} \perm(\theta)^M &= \left(\sum\limits_{\sigma \in {\mathcal S}_{m}}\theta_{1\sigma(1)} \theta_{2\sigma(2)}  \cdots \theta_{m\sigma(m)}\right)^M=
\sum_{|\vect{t_{j}}|=M} {M \choose {\vect{t_{j}}}} \prod\limits_{\sigma \in {\mathcal S}_{m}} \left(\theta_{1\sigma(1)} \theta_{2\sigma(2)}  \cdots \theta_{m\sigma(m)}\right)^{t_{j, \sigma}}.  
\end{align*} 
 \end{IEEEproof}
 In the next example, we illustrate the upper bound for the exponent matrix $R=R_\tau$ in Example \ref{example-lift-coefficient}.
      \begin{example}\label{coefficient} Let $M=3$, and let  $\theta$, $\theta ^{\uparrow\matr{P}}$, $R=R_\tau$ as in Example \ref{example-lift-coefficient} and $a^2bdf^2h^2i$  be the product corresponding to $R$. 
      How many permanent-products could exist in $\theta ^{\uparrow\matr{P}}$ that lead to the product $a^2bdf^2h^2i$?
      Note that $R$ has the unique standard decomposition $a^2bdf^2h^2i= (afh)^2(dbi)$. Therefore,  
  by Cor.~\ref{cor:coeff}, we expect no more than ${3\choose 2,1} =\frac{3!}{1!2!}=3$ permutations to result in this product.  
 Indeed,  there are exactly three combinations of same-index permanent-products in $\theta$ mapping into $(afh)^2(dbi)$, namely 
 $(a_1f_1h_1)(d_2b_2i_2)(a_3h_3f_3)$, $(a_1f_1h_1)(a_2h_2f_2)(d_3b_3i_3)$ and  $(d_1b_1i_1)(a_2f_2h_2)(a_3h_3f_3)$, giving  3  possible products of $M=3$ permanent-products in $\theta$
 with the standard decomposition $(afh)^2(dbi)$, i.e., a maximum of $N_{1}\defeq {3\choose 2,1} =3$ possible products.

Note that, in fact, all three above products of permanent products in $\theta$  are valid permanent-products in $\theta ^{\uparrow\matr{P}}$, resulting in the coefficient of $a^2bdf^2h^2i$ being equal to the upper bound 3. One of these products was     
      $A_\tau \defeq a_1d_2a_3h_1b_2h_3f_3f_1i_2=(a_1f_1h_1)(d_2b_2i_2)(a_3h_3f_3) = a^2bdf^2h^2i  $  of  Example~\ref{example-lift-coefficient}. 

Let us now compute the maximum coefficient of $ abcdefghi$ in $\theta ^{\uparrow\matr{P}}$. Observe that its exponent matrix has two possible standard decompositions: $(afh)(ceg)(bdi)$ and $ (aei)(bfg)(csh)$. 
Therefore, the maximum possible coefficient of $ abcdefghi$ in $\perm(\theta ^{\uparrow\matr{P}})$ is equal to the sum of two equal multinomial coefficients associated with the vector $(1,1,1)$, i.e., ${3\choose 1,1,1} + {3\choose 1,1,1}=2\frac{3!}{1!1!1!}=12.$ The actual coefficient 
of  $ abcdefghi$ in $\perm(\theta)^M$ is 0, which satisfies the upper bound trivially. 

In Example \ref{computation} in Appendix \ref{sec:app:perm}, we used Maple to compute the actual permanent of $\theta ^{\uparrow\matr{P}}$. We also expanded $(\perm(\theta))^3$ to illustrate the upper bound. 
 \end{example}

  \subsection{Bounding the degree M-Bethe and Bethe permanents}  \label{boundingBethe}
The bound $\perm(\theta ^{\uparrow\matr{\tilde P}})\leq \perm(\theta)^M$ gives  the following inequality conjectured  in~\cite{Vontobel:13} by applying the bound  to the permanent of each of the $M$-lifts of $\theta$, and hence also to their average, and then taking the $M$th root.
\begin{theorem} \label{M-Bethe}Let $\theta=(\theta_{ij})$ be a non-negative matrix of size $m\times m$  and $M\geq 1$ an integer.  
Then $$\perm_{\mathrm{B},M}(\theta) \leq \perm(\theta). $$
 \end{theorem} 
Taking the limit 
we obtain the following theorem.  
\begin{theorem} \label{BetheConjecture}Let $\theta=(\theta_{ij})$ be a non-negative matrix of size $m\times m$  and $M\geq 1$ an integer. Then $$\perm_{\mathrm{B}}(\theta)\leq  
\perm_{\mathrm{B},M}(\theta) \leq \perm(\theta). $$  \end{theorem} 
Note that the inequality $\perm_{\mathrm{B}}(\theta)\leq\perm(\theta)$ was proved by Gurvits in~\cite{2011arXiv1106.2844G} using a very different method. Our proof is a simple alternative  that uses only the combinatorial definition of the Bethe permanent.

\section{Conclusions} \label{conclusions}
 In this paper we proved two related conjectures posed by Vontobel in~\cite{Vontobel:13} on the permanent of an $M$-lift  $\theta^{\uparrow\matr{P}}$ of a matrix $\theta$ and on the degree $M$ Bethe permanent $\perm_{M,\mathrm{B}} (\theta)$ of $\theta$, namely, we show that $\perm(\theta^{\uparrow\matr{P}})\leq \perm(\theta)^M$ and, consequently,  
that $ \perm(\theta)$ of $\theta$, i.e.,  $\perm_{M, \mathrm{B}} (\theta)\leq  \perm(\theta)$.  As a corollary, our proof of these conjectures provides an alternative proof of  the inequality $\perm_{\mathrm{B}} (\theta)\leq  \perm(\theta)$ on the Bethe permanent of the base matrix $\theta$, one  that uses only the combinatorial Definition~\ref{Bethe} of the Bethe permanent from~\cite{Vontobel:13}. The first proof was given by Gurvits in~\cite{2011arXiv1106.2844G}.

The consequences of the results in this paper are more than just purely theoretical.  Apart from showing that it is possible to give a purely combinatorial proof that $\perm_{\mathrm{B}} (\theta)\leq  \perm(\theta)$ on the Bethe permanent of the base matrix $\theta$
(the earlier proof  \cite{2011arXiv1106.2844G} used different techniques), they provide new insight into the structure of the permanent of a $\matr{P}$-lifting of a matrix, which can be exploited algorithmically to decrease the computational complexity
of the permanent of the $\matr{P}$-liftings. Such an algorithm can search for products of groups of entries formed according to the decompositions presented in this paper to check if they form valid permanent-products.  

In addition, the structure of the permanent-products of  $\matr{P}$-liftings of a matrix may have some implications on the constant $C$ in the inequality $\perm(\theta) \leq C \cdot \perm_{\mathrm{B}} (\theta)$
in the conjectures stated by Gurvits in~\cite{2011arXiv1106.2844G}.
Lastly, since a $\matr{P}$-lifting of a matrix $\theta$ corresponds to an $M$-graph cover of the protograph (base graph)
described by $\theta$, which, in turn, correspond to LDPC codes, these results may help explain the
performance  of these codes through the techniques presented in\cite{MacKay:Davey:01:1} and extended and refined in~\cite{Smarandache:Vontobel:06:2:subm, Smarandache:Vontobel:09, 2012arXiv1201.2386B, 2011arXiv1111.0711W, 2012arXiv1210.3906P, Park:11} 
 for upper bounding the  minimum Hamming distance and the minimum pseudo-weight ~\cite{Wiberg:96} 
of a binary linear  code  that  is  described  by  an $m\times n$  parity-check  matrix $\matr{H}$.
This is done 
based on  explicitly constructing codewords and pseudo-codewords with components equal to determinants or permanents 
of some $m\times m$  submatrices  of $\matr{H}$ over the binary field or the ring of integers. 

\section*{Acknowledgment} 
We would like to thank Pascal O. Vontobel for suggesting this problem and for commenting on an earlier version of the paper. 

\appendices
\section{Examples of standard mapping}\label{sec:app:map}
In this appendix, we illustrate the standard mapping from Lemma \ref{big-lemma} by a few diverse examples.

\begin{example} Let $m=3$, $M=2$, and $\theta=(\theta_{ij})$  as in Example \ref{trivial}  and  let ${\theta}^{\uparrow\matr{P}}$ be defined as follows: 
\begin{align} \label{left-right}{\theta}^{\uparrow\matr{P}}\defeq \left[\begin{array}{cc|cc|cc} 
\shade{a_1}&0&b_1&0&c_1&0\\
0&\boxed{a_2}&0&b_2&0&c_2\\\hline
d_1&0&\shade{e_1}&0&f_1&0\\
0&d_2&0&e_2&0&\boxed{f_2}\\\hline
g_1&0&0&\shade{h_1}&0&i_1\\
0&g_2&h_2&0&\boxed{i_2}&0
\end{array}\right] = \left[\begin{array}{cc|cc|cc} 
\circled{$a_1$}&0&b_1&0&c_1&0\\
0&\boxed{\bf a_2}&0&b_2&0&c_2\\\hline
d_1&0&e_1&0&\circled{$f_1$}&0\\
0&d_2&0&\boxed{\bf e_2}&0&f_2\\\hline
g_1&0&0&h_1&0&\circled{$i_1$}\\
0&g_2&\boxed{\bf h_2}&0&i_2&0
\end{array} \right], 
\end{align} 
where in \eqref{left-right} (left matrix) we highlighted the permanent product $A_\tau\defeq (a_1e_1h_1)^\ill(a_2f_2i_2)^\ill=
(aei)(afh)$
and in \eqref{left-right} (right matrix) we highlighted the permanent product $A_\kappa\defeq (a_1f_1i_1)^\ill(a_2e_2h_2)^\ill=
(aei)(afh).$
These are both products of two illegal sub-products and have the same exponent matrix. 
In order to map these products, we need to list  the possible same-index decompositions for the two products: 
\begin{align*}A_{\tau,1}&\defeq (a_1e_1i_1)(a_2f_2h_2), \\ A_{\tau, 2}&\defeq (a_1f_1h_1) (a_2e_2i_2)\end{align*} 
and \begin{align*}
A_{\kappa,1}&\defeq (a_1e_1i_1)(a_2f_2h_2),\\A_{\kappa, 2}&\defeq (a_1f_1h_1) (a_2e_2i_2). 
\end{align*} 
Note that  $A_{\tau,i}= A_{\kappa,i}, $ for all $i=1,2$. 
Note also that the products are indexed from 1 to $M=2$ and that the entries in the sub-products are listed from top row to bottom row. 

Then $A_\tau \mapsto(a_1e_1i_1)(a_2f_2h_2) $  and $A_\kappa \mapsto (a_1f_1h_1) (a_2e_2i_2)$ because 
the first product $(a_1e_1h_1)$ of $A_\tau$ has its  first two entries equal to those of  $(a_1e_1i_1)$ and 
the first product $(a_1f_1i_1)$ of $A_\kappa$ has its  first two entries equal to those of  $(a_1f_1h_1)$. 
\end{example}

\begin{example}  \label{matrix5by3-legal}
Let $m=5$, $M=3$,  and  let 
{\small \begin{align*}
{\theta}^{\uparrow\matr{P}}\defeq &\setlength{\arraycolsep}{2.5pt}\left[\begin{array}{ccc|ccc|ccc|ccc|ccc}
a_{1}&0    &0    &d_{1}  &0  &0   &               c_{1}&0&0 &\boxed{d_{1}}&0&0& e_{1}&0&0\\ 
0&\circled{ $a_{2}$}&0    &0    &b_{2}  &0  &0   &               c_{2}&0&0 &d_{2}&0&0&                            e_{2}&0\\ 
0& 0&\shade{a_{3}}&0    & 0 &b_{3}  &0   &                     0&c_{3}&0&0&d_{3}&0&               0&e_{3}\\\hline
f_{1}&0    &0    &0  &0 &g_{1}  &         0&0 &     \boxed{h_{1}}&i_{1}&0&0&0& j_{1}&0\\ 
0&f_{2}&0        &g_{2}&0 &0  &0   &               h_{2}&0&0 &i_{2}&0&                             \circled{$j_{2}$}&0&0\\ 
0& 0& f_{3}&0   & g_{3}&0  &h_{3}&0   &                     0&0&0&     \shade{i_{3}}&0&               0&j_{3}\\\hline
k_{1}&0    &0    &0  &0 &\boxed{l_{1}}  &         0&      m_{1}&0& n_{1}&0&0&0& o_{1}&0\\ 
0&k_{2}&0        &\circled{ $l_{2}$}&0 &0  &0   &      0&        m_{2} &0 &n_{2}&0&                0&0&             o_{2}\\ 
0& 0&k_{3} &0   & \shade{l_{3}}&0  &m_{3}&0   &                     0&0&0&n_{3}&o_{3}&0&0\\\hline
\boxed{p_{1}}&0    &0    &0  &0 &q_{1} &              r_{1} &0&0& s_{1} &0&0&0&0& t_{1}\\ 
0&p_{2}&0        &q_{2}&0 &0  &0   &              \circled{$r_{2}$}&0& 0&s_{2}&0&                          t_{2} &0&0\\ 
0& 0& p_{3}&0   &q_{3}&0  &0&0&r_{3}&0&0&s_{3}&0&\shade{t_{3}}&0\\\hline
u_{1}&0    &0    &0  &0 &v_{1} &           0&0&    w_{1} &0&0&x_{1} &0&0& \boxed{y_{1} }\\ 
0&u_{2}&0        &v_{2}&0 &0  &       \circled{$ w_{2}$}    &0&0& x_{2}&0&0&                            y_{2}&0&0\\ 
0& 0& u_{3}&0   &v_{3}&0  &0& w_{3}&0&0&\shade{x_{3}} &0&0&y_{3}&0
\end{array}\right], 
\end{align*}}%
with the entries of the product 
$A_\tau\defeq a_{2}a_{3} d_{1} h_{1} i_{3} j_{2}  l_{1}  l_{2} l_{3 } p_{1}r_{2}t_{3}w_{ 2} x_{3}y_{ 1}=a^2  d h i j  l^3   prtw xy $ 
highlighted  in the matrix ${\theta}^{\uparrow\matr{P}}.$ All entries  
in a  $(i,j)$ block are equal to the entry $\theta_{ij}$, for all $ l\in [3]$, e.g., $a_1=a_2=a_3 =a$, where the index denotes the row position of that entry in $P_{ij}$, for example, $a_2$ is on the $2$nd row of $aP_{11}$. 
The following matrices can be computed: 
 \begin{align} \label{R-tau} R_\tau = \begin {bmatrix} 2&0&0&1&0\\0&0&1&1&1\\ 0&3&0&0&0\\ 1&0&1&0&1\\0&0&1&1&1\end {bmatrix}, \quad \overline{\alpha}_\tau =
\left[\begin{array}{c|c|c|c|c} \circled{$ 2$}~\shade{3}&\emptyset&\emptyset&\boxed{1}&\emptyset\\\hline
 \emptyset&\emptyset&\boxed{1}& \shade{3}& \circled{$ 2$}\\\hline
 \emptyset& \boxed{1}~\circled{$2$}~  \shade{3}& \emptyset&\emptyset&\emptyset\\\hline
  \boxed{1} &\emptyset&\circled{$2$}&\emptyset&\shade{3}\\ \hline
  \emptyset&\emptyset&\circled{$2$}&\shade{3}&\boxed{1}
 \end{array}\right],  
\end{align}
which gives the same-index decomposition
 $A_\tau=
 (p_1d_{1}   h_{1} l_{1}  y_{1})  
          (a_{2} j_{2} l_{2}  r_{2} w_{2})^\ill       
    (a_{3} i_{3} l_{3}  t_{3} x_{3})^\ill.$ 
    
    The following is the list of possible same-index products of permanent-products of $\theta$ that have the exponent matrix $R_\tau$: 
\begin{align*}
A_{\tau,1}\defeq (p_{1}d_{1}   h_{1} l_{1} y_{1}) (a_{2} j_{2} l_{2}  r_{2} x_{2}) (a_{3} i_{3}l_3 t_{3} w_{3}),\\
 A_{\tau,2}\defeq (p_{1}d_{1}   h_{1} l_{1} y_{1}) (a_{2} i_{2} l_{2}  t_{2} w_{2}) (a_{3} j_{3}l_3 r_{3} x_{3}),\\
A_{\tau,3} \defeq (p_{1}d_{1}  j_{1} l_{1}w_{1}) (a_{2} i_{2} l_{2}  r_{2}y _{2})(a_{3} h_{3} l_3t_{3} x_{3}),\\
A_{\tau,4}\defeq (p_{1}d_{1}  j_{1} l_{1}   w_{1})(a_{2} h_{2} l_{2}  t_{2} x_{2}) (a_{3} i_{3}l_3 r_{3} y_{3}). 
\end{align*}

The first term $(d_{1}   h_{1} l_{1}  p_{1} y_{1} )$  of $A_\tau$ is equal to the first term of $A_{\tau,1}$ and  $A_{\tau,2}$ and not equal to the first term of $A_{\tau,3}$ and $A_{\tau,3}$. The second term  $(a_{2} j_{2} l_{2}  r_{2} w_{2})^\ill$ has the first three terms equal to the first three terms of both  $(a_{2} j_{2} l_{2}  r_{2} x_{2})$ of   $A_{\tau,1}$ and $ (a_{2} i_{2} l_{2}  t_{2} w_{2})$ of $A_{\tau,2}$ and the forth term is equal to that of $(a_{2} j_{2} l_{2}  r_{2} x_{2})$ of    $A_{\tau,1}$ and not equal to that of $ (a_{2} i_{2} l_{2}  t_{2} w_{2})$ of $A_{\tau,2}$. 
Therefore, we map $A_\tau \mapsto A_{\tau,1}$. 
Note that $A_{\tau,1}$  cannot be a permanent-product in ${\theta}^{\uparrow\matr{P}}$ (otherwise  we would have a $2\times 2$ sub-matrix of  ${\theta}^{\uparrow\matr{P}}$ with $x_{2}$  and $x_{3}$ on the same column). 
\end{example} 
\clearpage
\begin{example} 
Let us now take $A_\gamma\defeq a_{1}a_{3} d_{2} h_{1} i_{3} j_{2}  l_{1}  l_{2} l_{3 } p_{2}r_{1}t_{3}w_{3} x_{2}y_{1}=a^2  d h i j  l^3   prtw xy$  illustrated through the highlighted entries in the matrix \par
{\small \begin{align*}{\theta}^{\uparrow\matr{P}}\defeq &\setlength{\arraycolsep}{2.5pt}\left[\begin{array}{ccc|ccc|ccc|ccc|ccc}
\boxed{a_{1}}&0    &0    &d_{1}  &0  &0   &               c_{1}&0&0 &{d_{1}}&0&0& e_{1}&0&0\\ 
0& a_{2}&0    &0    &b_{2}  &0  &0   &               c_{2}&0&0 &\circled{$d_{2}$}&0&0&                            e_{2}&0\\ 
0& 0&\shade{a_{3}}&0    & 0 &b_{3}  &0   &                     0&c_{3}&0&0&d_{3}&0&               0&e_{3}\\\hline
f_{1}&0    &0    &0  &0 &g_{1}  &         0&0 &     \boxed{h_{1}}&i_{1}&0&0&0& j_{1}&0\\ 
0&f_{2}&0        &g_{2}&0 &0  &0   &               h_{2}&0&0 &i_{2}&0&                             \circled{$j_{2}$}&0&0\\ 
0& 0& f_{3}&0   & g_{3}&0  &h_{3}&0   &                     0&0&0&     \shade{i_{3}}&0&               0&j_{3}\\\hline
k_{1}&0    &0    &0  &0 &\boxed{l_{1}}  &         0&      m_{1}&0& n_{1}&0&0&0& o_{1}&0\\ 
0&k_{2}&0        &\circled{ $l_{2}$}&0 &0  &0   &      0&        m_{2} &0 &n_{2}&0&                0&0&             o_{2}\\ 
0& 0&k_{3} &0   & \shade{l_{3}}&0  &m_{3}&0   &                     0&0&0&n_{3}&o_{3}&0&0\\\hline
p_{1}&0    &0    &0  &0 &q_{1} &              \boxed{r_{1}} &0&0& s_{1} &0&0&0&0& t_{1}\\
 0&\circled{$p_{2}$}&0        &q_{2}&0 &0  &0   &             r_{2}&0& 0&s_{2}&0&                          t_{2} &0&0\\
0& 0& p_{3}&0   &q_{3}&0  &0&0&r_{3}&0&0&s_{3}&0&\shade{t_{3}}&0\\\hline
u_{1}&0    &0    &0  &0 &v_{1} &           0&0&    w_{1} &0&0&x_{1} &0&0&\boxed{ y_{1}} \\ 
0&u_{2}&0        &v_{2}&0 &0  &       w_{2}    &0&0&  \circled{$x_{2}$}&0&0&                            y_{2}&0&0\\ 
0& 0& u_{3}&0   &v_{3}&0  &0& \shade{w_{3}}&0&0&x_{3} 
&0&0&y_{3}&0
\end{array}\right], 
\end{align*}} 
$A_\gamma$ has the same exponent matrix $R_\tau$ in \eqref{R-tau} and the index-matrix 
 \begin{align*} 
  \overline{\alpha}_\gamma =
\left[\begin{array}{c|c|c|c|c} \boxed{1}, \shade{3}&\emptyset&\emptyset&\circled{$ 2$}&\emptyset\\\hline
 \emptyset&\emptyset&\boxed{1}& \shade{3}& \circled{$ 2$}\\\hline
 \emptyset& \boxed{1},\circled{$2$},  \shade{3}& \emptyset&\emptyset&\emptyset\\\hline
  \circled{$2$} &\emptyset&\boxed{1}&\emptyset&\shade{3}\\ \hline
  \emptyset&\emptyset&\shade{3}&\circled{$2$}&\boxed{1}
 \end{array}\right],  
\end{align*}
which gives the same-index decomposition $A_\gamma=  (a_{1} h_{1} l_{1} r_1 y_{1})^\ill(p_2d_{2}   j_{2} l_{2}x_2) (a_{3} i_{3} l_{3}  t_{3} w_{3}).$
The  set associated with $A_\gamma$ is 
\begin{align*}
A_{\gamma,1}\defeq (a_{1} h_{1} l_{1}  t_{1} x_{1}) (p_{2}d_{2}   j_{2} l_{2}   w_{2}) (a_{3} i_{3}l_3 r_{3} y_{3}),\\
 A_{\gamma,2}\defeq (a_{1} i_{1} l_{1}  r_{1} y_{1}) (p_{2}d_{2}   j_{2} l_{2}   w_{2}) (a_{3} h_{3}l_3 x_{3} t_{3}),\\
 A_{\gamma,3}\defeq (a_{1} i_{1} l_{1}  t_{1} w_{1}) (p_{2}d_{2}   h_{2} l_{2}   y_{2}) (a_{3} j_{3}l_3 r_{3} x_{3}),\\
 A_{\gamma,4} \defeq(a_{1} j_{1} l_{1}  r_{1} x_{1}) (p_{2}d_{2}   h_{2} l_{2}   y_{2}) (a_{3} i_{3}l_3 t_{3} w_{3}).
\end{align*}
We see that $A_\gamma\mapsto A_{\gamma, 1}$ since the products $(a_{1} h_{1} l_{1} r_1 y_{1})$ and  $(a_{1} h_{1} l_{1}  t_{1} x_{1})$ have the first 3 entries in common. 
Note that $A_{\gamma,1 }$ is not a permanent-product.  
\end{example}

\begin{example} \label{matrix5by3} 
Let $m=5$, $M=3$, $\theta$ and  ${\theta}^{\uparrow\matr{P}}$ be as in Example~\ref{matrix5by3-legal}.  
Let 
$A_\kappa\defeq   p_{1}a_{2}u_{3} g_{2}  l_{3}  d_{3}    e_{1} g_{1} h_{3} n_{2} o_{1}r_{2}r_{3}x_{2}y_{1}= (p_1e_{1}g_{1} o_{1}y_{1})^\ill(a_{2}g_{2}n_{2} r_{2} x_{2})^\ill 
(u_{3} d_{3} h_{3}l_{3}r_{3})^\ill= a d  e g^2 hl   n opr^2u  xy$  
as highlighted below, together with its exponent matrix and its index matrix.  \par
{\footnotesize \begin{align*}{\theta}^{\uparrow\matr{P}}= &\setlength{\arraycolsep}{2.5pt}\left[\begin{array}{ccc|ccc|ccc|ccc|ccc}
a_{1}&0    &0    &b_{1}  &0  &0   &               c_{1}&0&0 &d_{1}&0&0&\shade{ e_{1}}&0&0\\ 
0&\circled{ $a_{2}$}&0    &0    &b_{2}  &0  &0   &               c_{2}&0&0 &d_{2}&0&0&                            e_{2}&0\\ 
0& 0&a_{3}&0    & 0 &b_{3}  &0   &                     0&c_{3}&0&0&\boxed{d_{3}}&0&               0&e_{3}\\\hline
f_{1}&0    &0    &0  &0 &\shade{g_{1}} &         0&0 &     h_{1}&i_{1}&0&0&0& j_{1}&0\\ 
0&f_{2}&0        &\circled{$g_{2}$}&0 &0  &0   &               h_{2}&0&0 &i_{2}&0&                            {j_{2}}&0&0\\ 
0& 0& f_{3}&0   & g_{3}&0  &\boxed{h_{3}}&0   &                     0&0&0&     i_{3}&0&               0&j_{3}\\\hline
k_{1}&0    &0    &0  &0 &l_{1}  &         0&      m_{1}&0& n_{1}&0&0&0&\shade{o_{1}}&0\\ 
0&k_{2}&0        &l_{2}&0 &0  &0   &      0&        m_{2} &0 &\circled{$n_{2}$}&0&                0&0&             o_{2}\\ 
0& 0&k_{3} &0   & \boxed{l_{3}}&0  &m_{3}&0   &                     0&0&0&n_{3}&o_{3}&0&0\\\hline
\shade{p_{1}}&0    &0    &0  &0 &q_{1} &              r_{1} &0&0& s_{1} &0&0&0&0& t_{1}\\ 
0&p_{2}&0        &q_{2}&0 &0  &0   &              \circled{$r_{2}$}&0& 0&s_{2}&0&                          t_{2} &0&0\\ 
0& 0& p_{3}&0   &q_{3}&0  &0&0&\boxed{r_{3}}&0&0&s_{3}&0&t_{3}&0\\\hline
u_{1}&0    &0    &0  &0 &v_{1} &           0&0&    w_{1} &0&0&x_{1} &0&0& \shade{y_{1} }\\ 
0&u_{2}&0        &v_{2}&0 &0  &           w_{2}  
&0&0& \circled{$x_{2}$}&0&0&                            y_{2}&0&0\\ 
0& 0& \boxed{u_{3}}&0   &v_{3}&0  &0& w_{3}&0&0&x_{3}
&0&0&y_{3}&0
\end{array}\right]
\end{align*}}
\begin{align}\label{matrixRkappa}  R_\kappa = \begin {bmatrix} 1&0&0&1&1\\0&2&1&0&0\\ 0&1&0&1&1\\ 1&0&2&0&0\\1&0&0&1&1\end {bmatrix},  \quad \overline{\alpha}_\kappa =
\left[\begin{array}{c|c|c|c|c} \circled{$ 2$}&\emptyset&\emptyset&\boxed{3}&\shade{1}\\\hline
 \emptyset&\shade{1}~ \circled{$ 2$}&\boxed{3}&\emptyset&\emptyset \\\hline
 \emptyset& \boxed{3} & \emptyset&\circled{$2$}& \shade{1}\\\hline
 \shade{1} &\emptyset&\circled{$2$}~\boxed{3} &\emptyset&\emptyset\\ \hline
 \boxed{3} & \emptyset&\emptyset&\circled{$2$}&\shade{1}
 \end{array}\right],  
\end{align}
The set associated with $A_\kappa$ is 
\begin{align*}A_{\kappa, 1}\defeq (p_1d_1h_1l_1y_1)(a_{2}g_{2}o_{2} r_{2} x_{2})(u_{3} e_{3} g_{3}n_{3}r_{3}), \\
 A_{\kappa, 2}\defeq (p_1e_1h_1l_1x_1)(a_{2}g_{2}n_{2} r_{2} y_{2}) (u_{3} d_{3} g_{3}o_{3}r_{3}). \end{align*}
We see that $A_\kappa \mapsto A_{\kappa, 2}$ since the products $(p_1e_{1}g_{1} o_{1}y_{1})$ and $(p_1e_1h_1l_1x_1)$ and have the first 2 entries in common while $(p_1e_{1}g_{1} o_{1}y_{1})$ and $(p_1d_1h_1l_1y_1)$ do not. Note that $A_{\kappa, 2}$ is
not a permanent-product. 
\end{example}
\begin{example} Let $m=5$, $M=3$, $\theta$ and  
{\footnotesize \begin{align}
{\theta}^{\uparrow\matr{P}}\defeq &\setlength{\arraycolsep}{2.5pt}\left[\begin{array}{ccc|ccc|ccc|ccc|ccc}
a_{1}&0    &0    &b_{1}  &0  &0   &               c_{1}&0&0 &\shade{d_{1}}&0&0&{ e_{1}}&0&0\\ 
0&\circled{ $a_{2}$}&0    &0    &b_{2}  &0  &0   &               c_{2}&0&0 &d_{2}&0&0&                            e_{2}&0\\ 
0& 0&a_{3}&0    & 0 &b_{3}  &0   &                     0&c_{3}&0&0&{d_{3}}&0&               0&\boxed{e_{3}}\\\hline
f_{1}&0    &0    &0  &0 &\shade{g_{1}} &         0&0 &     h_{1}&i_{1}&0&0&0& j_{1}&0\\ 
0&f_{2}&0        &\circled{$g_{2}$}&0 &0  &0   &               h_{2}&0&0 &i_{2}&0&                            {j_{2}}&0&0\\ 
0& 0& f_{3}&0   & g_{3}&0  &\boxed{h_{3}}&0   &                     0&0&0&     i_{3}&0&               0&j_{3}\\\hline
k_{1}&0    &0    &0  &0 &l_{1}  &         0&      m_{1}&0& n_{1}&0&0&0&\shade{o_{1}}&0\\ 
0&k_{2}&0        &l_{2}&0 &0  &0   &      0&        m_{2} &0 &\circled{$n_{2}$}&0&                0&0&             o_{2}\\ 
0& 0&k_{3} &0   & \boxed{l_{3}}&0  &m_{3}&0   &                     0&0&0&n_{3}&o_{3}&0&0\\\hline
\shade{p_{1}}&0    &0    &0  &0 &q_{1} &              r_{1} &0&0& s_{1} &0&0&0&0& t_{1}\\ 
0&p_{2}&0        &q_{2}&0 &0  &0   &              \circled{$r_{2}$}&0& 0&s_{2}&0&                          t_{2} &0&0\\ 
0& 0& p_{3}&0   &q_{3}&0  &0&0&\boxed{r_{3}}&0&0&s_{3}&0&t_{3}&0\\\hline
u_{1}&0    &0    &0  &0 &v_{1} &           0&0&    w_{1} &0&0&\shade{x_{1}} &0&0& {y_{1} }\\ 
0&u_{2}&0        &v_{2}&0 &0  &           w_{2}   
&0&0& x_{2}&0&0&                            \circled{$y_{2}$}&0&0\\ 
0& 0& \boxed{u_{3}}&0   &v_{3}&0  &0& w_{3}&0&0&x_{3}&0&0&y_{3}&0
\end{array}\right] \label{theta_p2}
\end{align}}
(as in Example~\ref{matrix5by3-legal}).
Lastly, let  $$A_\nu=p_{1}a_{2}u_{3} g_{2}  l_{3}  d_{3}    e_{1} g_{1} h_{3} n_{2} o_{1}r_{2}r_{3}x_{2}y_{1} =
 (p_1 d_{1}g_{1}o_{1}x_{1})^\ill(a_{2}g_{2}n_{2} r_{2} y_{2}) 
(u_{3} e_{3} h_{3}l_{3}r_{3})^\ill
 = a d  e g^2 hl   n opr^2u  xy,$$
which is the product highlighted in \eqref{theta_p2}.

The exponent matrix of $A_\nu$ is  $R_\nu=R_\kappa$, its  index matrix is
\begin{align}\label{matrixRnu} 
\overline{\alpha}_\nu =
\left[\begin{array}{c|c|c|c|c} \circled{$ 2$}&\emptyset&\emptyset&\shade{1}&\boxed{3}\\\hline
 \emptyset&\shade{1}~ \circled{$ 2$}&\boxed{3}&\emptyset&\emptyset \\\hline
 \emptyset& \boxed{3} & \emptyset&\circled{$2$}& \shade{1}\\\hline
 \shade{1} &\emptyset&\circled{$2$}~\boxed{3} &\emptyset&\emptyset\\ \hline
 \boxed{3} & \emptyset&\emptyset&\shade{1}&\circled{$2$}
 \end{array}\right],  
\end{align}
and the associated set of products $A_\nu$ is 
\begin{align*} A_{\nu,1}\defeq  (p_1d_1h_1l_1y_1)(a_{2}g_{2}o_{2} r_{2} x_{2})(u_{3} e_{3} g_{3}n_{3}r_{3})=A_{\kappa, 1}, \\
 A_{\nu,1}\defeq (p_1e_1h_1l_1x_1)(a_{2}g_{2}n_{2} r_{2} y_{2}) (u_{3} d_{3} g_{3}o_{3}r_{3})=A_{\kappa, 2}. \end{align*}
We see that $A_\nu\mapsto A_{\kappa, 1}$ since the products $(p_1 d_{1}g_{1}o_{1}x_{1})$ and  $(p_1d_1h_1l_1y_1)$ have the first 2 entries in common while $(p_1 d_{1}g_{1}o_{1}x_{1})$ and $ (p_1e_1h_1l_1x_1)$ do not. Note that $A_{\kappa, 1}$ is not a permanent-product.  
\end{example} 

\section{Example for permanent bounds}\label{sec:app:perm}
Here we give an example comparing $\perm({\theta}^{\uparrow\matr{P}})$ with 
$(\perm(\theta))^3$.
 \begin{example}\label{computation}  Computing $\perm({\theta}^{\uparrow\matr{P}})$ we get
 \begin{align}\label{permanent}\nonumber\perm({\theta}^{\uparrow\matr{P}})&={a}^{3}{e}^{3}{i}^{3}+{a}^{3}{f}^{3}{h}^{3}+3\,{a}^{2}bd{f}^{2}{h}^{2}
i+3\,{a}^{2}b{f}^{3}g{h}^{2}+3\,{a}^{2}cd{e}^{2}h{i}^{2}+3\,{a}^{2}c{e
}^{2}fghi+3\,a{b}^{2}{d}^{2}fh{i}^{2}\\&+ 6\,a{b}^{2}d{f}^{2}ghi+3\,a{b}^{
2}{f}^{3}{g}^{2}h+3\,abcd{e}^{2}g{i}^{2}+3\,abc{e}^{2}f{g}^{2}i+3\,a{c
}^{2}{d}^{2}e{h}^{2}i+3\,a{c}^{2}defg{h}^{2}+{b}^{3}{d}^{3}{i}^{3}\nonumber\\&+3\,
{b}^{3}{d}^{2}fg{i}^{2}+3\,{b}^{3}d{f}^{2}{g}^{2}i+{b}^{3}{f}^{3}{g}^{
3}+3\,b{c}^{2}{d}^{2}eghi+3\,b{c}^{2}def{g}^{2}h+{c}^{3}{d}^{3}{h}^{3}
+{c}^{3}{e}^{3}{g}^{3}.
\end{align}
Computing $(\perm(\theta))^3$,  we can easily verify that all the products in  $\perm({\theta}^{\uparrow\matr{P}})$ appear in $(\perm(\theta))^3$ with a larger or equal coefficient, as predicted by Theorem~\ref{the-theorem}:
\begin{align*} (\perm(\theta))^3&=(aei+afh+bdi+bfg+cdh+egc)^3= {{\it aei}}^{3}+3\,{{\it aei}}^{2}{\it afh}+3\,{{\it aei}}^{2}{\it bdi
}+3\,{{\it aei}}^{2}{\it bfg}+3\,{{\it aei}}^{2}{\it cdh}
\\&
+3\,{{\it aei}}^{2}{\it egc}+3\,{\it aei}\,{{\it afh}}^{2}+6\,{\it aei}\,{\it afh}
\,{\it bdi}+6\,{\it aei}\,{\it afh}\,{\it bfg}+6\,{\it aei}\,{\it afh}
\,{\it cdh}+6\,{\it aei}\,{\it afh}\,{\it egc}
\\&
+3\,{\it aei}\,{{\it bdi}}^{2}+6\,{\it aei}\,{\it bdi}\,{\it bfg}+6\,{\it aei}\,{\it bdi}\,{
\it cdh}+6\,{\it aei}\,{\it bdi}\,{\it egc}+3\,{\it aei}\,{{\it bfg}}^
{2}+6\,{\it aei}\,{\it bfg}\,{\it cdh}+6\,{\it aei}\,{\it bfg}\,{\it 
egc}
\\&
+3\,{\it aei}\,{{\it cdh}}^{2}+6\,{\it aei}\,{\it cdh}\,{\it egc}+
3\,{\it aei}\,{{\it egc}}^{2}+{{\it afh}}^{3}+3\,{{\it afh}}^{2}{\it 
bdi}+3\,{{\it afh}}^{2}{\it bfg}+3\,{{\it afh}}^{2}{\it cdh}+3\,{{\it 
afh}}^{2}{\it egc}
\\&
+3\,{\it afh}\,{{\it bdi}}^{2}+6\,{\it afh}\,{\it 
bdi}\,{\it bfg}+6\,{\it afh}\,{\it bdi}\,{\it cdh}+6\,{\it afh}\,{\it 
bdi}\,{\it egc}+3\,{\it afh}\,{{\it bfg}}^{2}+\!6\,{\it afh}\,{\it bfg}
\,{\it cdh}+\!6\,{\it afh}\,{\it bfg}\,{\it egc}
\\&
+3\,{\it afh}\,{{\it cdh}}^{2}+6\,{\it afh}\,{\it cdh}\,{\it egc}+3\,{\it afh}\,{{\it egc}}^{2
}+{{\it bdi}}^{3}+3\,{{\it bdi}}^{2}{\it bfg}+3\,{{\it bdi}}^{2}{\it 
cdh}+3\,{{\it bdi}}^{2}{\it egc}+3\,{\it bdi}\,{{\it bfg}}^{2}
\\&
+6\,{\it bdi}\,{\it bfg}\,{\it cdh}+6\,{\it bdi}\,{\it bfg}\,{\it egc}+3\,{
\it bdi}\,{{\it cdh}}^{2}+6\,{\it bdi}\,{\it cdh}\,{\it egc}+3\,{\it 
bdi}\,{{\it egc}}^{2}+{{\it bfg}}^{3}+3\,{{\it bfg}}^{2}{\it cdh}
\\&
+3\,{
{\it bfg}}^{2}{\it egc}+3\,{\it bfg}\,{{\it cdh}}^{2}
+6\,{\it bfg}\,{\it cdh}\,{\it egc}+3\,{\it bfg}\,{{\it egc}}^{2}+{{\it cdh}}^{3}+3\,{
{\it cdh}}^{2}{\it egc}+3\,{\it cdh}\,{{\it egc}}^{2}+{{\it egc}}^{3}.
\end{align*}
\vspace*{-0.4cm}
\par
 \end{example} 
 
\bibliographystyle{ieeetr}

 \end{document}